\documentclass[
  journal=pasa,
  manuscript=research-paper, 
  year=202X,
  volume=YY,
]{cup-journal}

\usepackage{bm}
\usepackage{amsmath}
\usepackage{amssymb,microtype,siunitx,booktabs}
\usepackage{hyperref}
\usepackage{orcidlink} 
\hypersetup{colorlinks=true,linkcolor=blue,urlcolor=blue,linktoc=all,citecolor=blue}
\newcommand{\EWha}{${\rm EW}_{{\rm H}\alpha}$}
\newcommand{\lEWha}{${\log(\rm EW}_{{\rm H}\alpha})$}

\title{Hector Galaxy Survey: Linking the low- and high-mass ends of the initial mass function in star-forming galaxies}

\author{\orcidlink{https://orcid.org/0009-0002-2841-7038} D. Salvador}
\affiliation{School of Mathematical and Physical Sciences, Macquarie University, Sydney, NSW 2109, Australia}
\alsoaffiliation{Astrophysics and Space Technologies Research Centre, Macquarie University, Sydney, NSW 2109, Australia}
\email[D. Salvador]{diego.salvador@hdr.mq.edu.au}

\author{\orcidlink{https://orcid.org/0000-0002-6097-2747} A. M. Hopkins}
\affiliation{School of Mathematical and Physical Sciences, Macquarie University, Sydney, NSW 2109, Australia}
\alsoaffiliation{Astrophysics and Space Technologies Research Centre, Macquarie University, Sydney, NSW 2109, Australia}

\author{M. S. Owers}
\affiliation{School of Mathematical and Physical Sciences, Macquarie University, Sydney, NSW 2109, Australia}
\alsoaffiliation{Astrophysics and Space Technologies Research Centre, Macquarie University, Sydney, NSW 2109, Australia}
\alsoaffiliation{ARC Centre of Excellence for All Sky Astrophysics in 3 Dimensions (ASTRO 3D), Australia}

\author{\orcidlink{https://orcid.org/0000-0001-8494-0080} I. Martín-Navarro}
\affiliation{Instituto de Astrof\'{\i}sica de Canarias,c/ V\'{\i}a L\'actea s/n, E38205 - La Laguna, Tenerife, Spain}
\alsoaffiliation{Departamento de Astrof\'isica, Universidad de La Laguna, E-38205 La Laguna, Tenerife, Spain}

\author{\orcidlink{0009-0009-9074-716X} G. Quattropani}
\affiliation{School of Mathematical and Physical Sciences, Macquarie University, Sydney, NSW 2109, Australia}
\alsoaffiliation{Astrophysics and Space Technologies Research Centre, Macquarie University, Sydney, NSW 2109, Australia}
\alsoaffiliation{ARC Centre of Excellence for All Sky Astrophysics in 3 Dimensions (ASTRO 3D), Australia}

\author{\orcidlink{0000-0002-4326-8598} P. K. Das}
\affiliation{School of Mathematics and Physics, University of Queensland, Brisbane, QLD 4072, Australia}

\author{\orcidlink{0000-0002-5896-0034} M. Pak}
\affiliation{School of Mathematical and Physical Sciences, Macquarie University, Sydney, NSW 2109, Australia}
\alsoaffiliation{ARC Centre of Excellence for All Sky Astrophysics in 3 Dimensions (ASTRO 3D), Australia}
\alsoaffiliation{Korea Astronomy and Space Science Institute (KASI), 776 Daedeok-daero, Yuseong-gu, Daejeon  34055, Republic of Korea}

\author{\orcidlink{0000-0003-2880-9197} S. M. Croom}
\affiliation{Sydney Institute for Astronomy (SIfA), School of Physics, The University of Sydney, Sydney, NSW 2006, Australia}
\alsoaffiliation{ARC Centre of Excellence for All Sky Astrophysics in 3 Dimensions (ASTRO 3D), Australia}

\author{J. J. Bryant}
\affiliation{Sydney Institute for Astronomy (SIfA), School of Physics, The University of Sydney, Sydney, NSW 2006, Australia}
\alsoaffiliation{Astralis-USyd, Sydney Institute for Astronomy, School of Physics, The University of Sydney, Sydney, NSW 2006, Australia}

\author{\orcidlink{https://orcid.org/0000-0002-1251-9905} T. Jeřábková}
\affiliation{Department of Theoretical Physics and Astrophysics, Faculty of Science, Masaryk University, Kotlářská 2, Brno 611 37, Czech Republic}

\author{\orcidlink{0000-0002-3254-9044} K. Glazebrook}
\affiliation{Centre for Astrophysics and Supercomputing, Swinburne University of Technology, PO Box 218, Hawthorn, VIC 3122, Australia}

\author{\orcidlink{0000-0003-2723-0810} A. Ristea}
\affiliation{Centre for Astrophysics and Supercomputing, Swinburne University of Technology, PO Box 218, Hawthorn, VIC 3122, Australia}
\alsoaffiliation{ARC Centre of Excellence in Optical Microcombs for Breakthrough Science (COMBS)}

\author{\orcidlink{0000-0002-7301-461X} M. L. P. Gunawardhana}
\affiliation{Sydney Institute for Astronomy (SIfA), School of Physics, The University of Sydney, Sydney, NSW 2006, Australia}
\alsoaffiliation{Research School of Astronomy and Astrophysics, Australian National University, Canberra, ACT 2611, Australia}
\alsoaffiliation{ARC Centre of Excellence for All Sky Astrophysics in 3 Dimensions (ASTRO 3D), Australia}

\author{\orcidlink{0000-0002-1576-2505} S. M. Sweet}
\affiliation{School of Mathematics and Physics, University of Queensland, Brisbane, QLD 4072, Australia}
\alsoaffiliation{ARC Centre of Excellence for All Sky Astrophysics in 3 Dimensions (ASTRO 3D), Australia}

\author{\orcidlink{0000-0002-5037-951X} K. Oh}
\affiliation{Korea Astronomy and Space Science Institute (KASI), 776 Daedeok-daero, Yuseong-gu, Daejeon  34055, Republic of Korea}

\author{J. C. Lee}
\affiliation{Korea Astronomy and Space Science Institute (KASI), 776 Daedeok-daero, Yuseong-gu, Daejeon  34055, Republic of Korea}

\author{\orcidlink{0000-0003-3451-0925} J. H. Lee}
\affiliation{Korea Astronomy and Space Science Institute (KASI), 776 Daedeok-daero, Yuseong-gu, Daejeon  34055, Republic of Korea}

\author{\orcidlink{https://orcid.org/0000-0003-0247-1204} C. Foster}
\affiliation{School of Physics, University of New South Wales, Sydney, NSW 2052, Australia}
\alsoaffiliation{ARC Centre of Excellence for All Sky Astrophysics in 3 Dimensions (ASTRO 3D), Australia}

\author{T. Zafar}
\affiliation{School of Mathematical and Physical Sciences, Macquarie University, Sydney, NSW 2109, Australia}
\alsoaffiliation{Astrophysics and Space Technologies Research Centre, Macquarie University, Sydney, NSW 2109, Australia}

\author{\orcidlink{0000-0003-3514-6280} Y. Mai}
\affiliation{Australian Astronomical Optics, Macquarie University, Sydney, NSW 2109, Australia}
\alsoaffiliation{Astrophysics and Space Technologies Research Centre, Macquarie University, Sydney, NSW 2109, Australia}
\alsoaffiliation{ARC Centre of Excellence for All Sky Astrophysics in 3 Dimensions (ASTRO 3D), Australia}

\author{\orcidlink{0009-0002-8534-5077} S. Tuntipong}
\affiliation{Sydney Institute for Astronomy (SIfA), School of Physics, The University of Sydney, Sydney, NSW 2006, Australia}
\alsoaffiliation{ARC Centre of Excellence for All Sky Astrophysics in 3 Dimensions (ASTRO 3D), Australia}

\author{\orcidlink{0000-0003-2552-0021} J. van de Sande}
\affiliation{School of Physics, University of New South Wales, Sydney, NSW 2052, Australia}
\alsoaffiliation{ARC Centre of Excellence for All Sky Astrophysics in 3 Dimensions (ASTRO 3D), Australia}

\author{S. Barsanti}
\affiliation{Sydney Institute for Astronomy (SIfA), School of Physics, The University of Sydney, Sydney, NSW 2006, Australia}
\alsoaffiliation{Research School of Astronomy and Astrophysics, Australian National University, Canberra, ACT 2611, Australia}

\author{\orcidlink{0000-0001-7516-4016} J. Bland-Hawthorn}
\affiliation{Sydney Institute for Astronomy (SIfA), School of Physics, The University of Sydney, Sydney, NSW 2006, Australia}
\alsoaffiliation{ARC Centre of Excellence for All Sky Astrophysics in 3 Dimensions (ASTRO 3D), Australia}

\author{\orcidlink{0000-0001-9552-8075} M. Colless}
\affiliation{Research School of Astronomy and Astrophysics, Australian National University, Canberra, ACT 2611, Australia}

\author{R. Content}
\affiliation{Astralis-AAO, Australian Astronomical Optics, Faculty of Science and Engineering, Macquarie University, NSW 2109, Australia}

\author{T. Farrell}
\affiliation{Astralis-AAO, Australian Astronomical Optics, Faculty of Science and Engineering, Macquarie University, NSW 2109, Australia}

\author{J. Lawrence}
\affiliation{Astralis-AAO, Australian Astronomical Optics, Faculty of Science and Engineering, Macquarie University, NSW 2109, Australia}

\author{S. Min}
\affiliation{Astralis-USyd, Sydney Institute for Astronomy, School of Physics, The University of Sydney, Sydney, NSW 2006, Australia}

\author{\orcidlink{0000-0002-4731-9604} S. Oh}
\affiliation{Department of Astronomy and Yonsei University Observatory, Yonsei University, Seoul, 03722, Republic of Korea}

\author{N. Pai}
\affiliation{Astralis-AAO, Australian Astronomical Optics, Faculty of Science and Engineering, Macquarie University, NSW 2109, Australia}

\author{A. Salim Sadman}
\affiliation{Business Systems Analyst, The University of New South Wales, Sydney, Australia}

\author{R. Zhelem}
\affiliation{Astralis-AAO, Australian Astronomical Optics, Faculty of Science and Engineering, Macquarie University, NSW 2109, Australia}

\received {dd Mmm YYYY}
\revised  {dd Mmm YYYY}
\accepted {dd Mmm YYYY}
\published{DD MMMMMMMMM 2025}


\begin{document}


\begin{abstract} 
The stellar initial mass function (IMF) is a fundamental ingredient in galaxy evolution, linking observed integrated light to stellar masses, star-formation rates, and chemical enrichment histories. Constraining the full IMF shape beyond the Milky Way remains challenging, as most studies focus either on the low-mass end of quiescent galaxies or the high-mass end of star-forming galaxies. Here we present the first simultaneous analysis of both ends of the stellar initial mass function (IMF) in 214 star-forming galaxies from the Hector survey (z $\sim$ 0.01–0.07). We estimate the low-mass end slope ($\alpha_{\rm low}$) using a stellar population approach that fits IMF-sensitive absorption features with extended star formation histories, while the high-mass end slope ($\alpha_{\rm high}$) is derived via the Kennicutt diagnostic, which compares the observed H$\alpha$ equivalent width and $g-r$ colour with stellar population synthesis models. We find substantial diversity in IMF shapes, with galaxies spanning combinations of bottom-heavy/light and top-heavy/light slopes. A weak but statistically robust correlation between the low- and high-mass IMF slopes is observed, but partial correlation analysis indicates that this apparent link is largely driven by their mutual dependence on stellar mass and metallicity. Both IMF slopes show significant correlations with stellar mass, star formation activity (traced by H$\alpha$ luminosity and surface density), and stellar metallicity ([M/H]). In general, higher stellar mass, stronger star formation activity, and higher metallicity are associated with both bottom-heavy and top-heavy IMFs. We find that the full IMF shape seems to be modulated by total stellar mass. Partial correlation analysis reveals that $\alpha_{\rm low}$ is primarily driven by [M/H], whereas $\alpha_{\rm high}$ is mainly linked to stellar mass and recent star formation. Because $\alpha_{\rm low}$ traces the IMF over long-term averages and $\alpha_{\rm high}$ captures only recent ($\lesssim 10$ Myr) star formation, the processes shaping each end likely occur over different and possibly decoupled timescales. Our findings challenge the universality of the IMF and emphasise the need for galaxy evolution and stellar population models to incorporate a flexible IMF prescription. Accounting for these variations is essential to build an IMF-consistent picture of galaxy evolution across cosmic time.
\end{abstract}

\section{INTRODUCTION}
\label{sec:int}


The physical properties of galaxies, and their evolution, are fundamentally shaped by the type and amount of stars they form throughout their history. Observers infer these properties from the light emitted by unresolved stellar populations, particularly when studying distant galaxies. In this context, the initial mass function (IMF), the theoretical distribution of stellar masses produced in a star formation event, plays a crucial role in inferring these properties. The IMF sets the relative number of stars of different masses formed in such an event and, as a result, determines the fraction of light contributed by each mass range to the total light budget. This makes the IMF a key input assumption in models that aim to reproduce the integrated light spectra or photometric properties of galaxies \citep{Walcher_2010,ConroyARAA_2013}, to link observed luminosities with stellar masses \citep{Chabrier_2003}, and to predict their star formation and chemical enrichment histories \citep[e.g.,][]{Ferreras_2015,Gutcke_2019,navarro_2021}.

Early IMF studies within the solar neighbourhood and the Milky Way \citep[e.g.,][]{salpeter_1955,millerscalo_1979} showed that the IMF is well described by a power law with slope $\alpha \sim -2.35$ (or $\Gamma \sim -1.35$ in logarithmic notation) for stars more massive than $\sim \rm 0.5$–$\rm 1\,M_\odot$. The IMF slope characterises the number of stars per mass interval in each segment of the piecewise function commonly used to describe the IMF \citep{kroupa_2001,Chabrier_2003}. Typically, the mass range from $\sim0.08\rm \,M_\odot$ to $\sim0.5$–$1\, \rm M_\odot$ is referred to as the low-mass end of the IMF, while the high-mass end spans from $\sim0.5$–$1\rm \,M_\odot$ up to $\sim120$–$150\rm \,M_\odot$. Some authors further divide the IMF into three regimes, separating the brown dwarf/sub-stellar domain ($M \lesssim 0.08\, \rm\, M_\odot$), the stellar low-mass regime ($\sim0.08$–$0.5\ \rm\, M_\odot$), and the intermediate-to-high mass regime ($M \gtrsim 0.5 \, \rm M_\odot$; e.g., \citealt{kroupa_2001}; \citealt{Chabrier_2003}).

In nearby galaxies within the Local Group ($z \lesssim 0.001$), where individual stars over a wide mass range can be resolved, the IMF has been directly estimated through stellar counts, yielding results consistent with those in the solar neighbourhood \citep[e.g.,][]{MasseyHunter_1998,Massey_2002,geha_2013}. The consistency of these results across environments with different stellar densities (e.g., \citealt{Massey_1998,kroupa_2001}; see also the review by \citealt{bastian_2010}) established the widely adopted assumption of an environment-invariant IMF. 

In distant galaxies ($z \gtrsim 0.05$), where individual stars cannot be resolved, the IMF can still be constrained through three main observational approaches. A first approach is based on dynamical or gravitational methods. As low-mass stars make up most of a galaxy’s stellar mass and high-mass stars dominate its luminosity, the mass-to-light ratio (\rm $M/L$) serves as a proxy for the relative abundance of low- to high-mass stars, and hence for the IMF. Therefore, once the luminosity of a stellar population is measured, its stellar mass can be independently estimated using dynamical constraining techniques \citep[e.g., via integral-field stellar kinematics,][]{cappellari_2012} or via gravitational lensing analyses \citep{auger_2010,treu_2010,leier_2016}. A second approach, more recently developed, involves the use of chemical abundance models based on isotopic ratios, such as $^{12}$C/$^{13}$C, $^{16}$O/$^{18}$O and $^{14}$N/$^{15}$N \citep[e.g.,][]{Sliwa_2017,Romano_2019,Guo_2024}. These are measured in the submillimetre/millimetre regimes—wavelengths less affected by dust extinction. These models take advantage of the fact that different isotopes are synthesised in stars of varying masses and are released into the interstellar medium on different timescales. While promising, this method depends on assumptions regarding stellar yields and the chemical evolution histories of galaxies. 

A third approach uses the integrated light spectrum of unresolved stellar populations, which encodes the contribution of stars across the full mass range. The shape of the IMF can then be constrained by analysing spectral features sensitive to different stellar mass regimes. The high-mass end is typically probed through emission lines arising from ionised gas surrounding massive stars, such as H$\alpha$ \citep[e.g.,][]{hoversten_2008,hoversten_2010,gunawardhana_2011,nanayakkara_2017,Salvador_2025}, or the ratio of H$\alpha$ to UV emission \citep[][]{meurer_2009,lee_2009}. Conversely, the low-mass end is constrained from absorption features whose strengths depend on the stellar surface gravity, which reflects the relative contribution of low-mass stars (mainly M-dwarfs) to the stellar population. These include the Wing-Ford band \citep[e.g.,][]{vanDokkum_2010,vanDokkum_2012,conroy_2012}, the NaI doublet \citep[e.g.,][]{Labarbera_2013}, and the CaH and TiO molecular bands \citep[e.g.,][]{spiniello_2014}.

It is worth noting that, although dynamical and stellar population-based methods often yield quantitatively different results \citep[][]{smith_2014}, they consistently reveal similar trends with galaxy mass and radial gradients. This convergence lends strong support to the idea of IMF non-universality, at least in early-type galaxies (ETGs).

To date, no single method has yet been able to estimate the IMF simultaneously across the full stellar mass range for a particular type of stellar population (e.g., star-forming and quiescent galaxies). Instead, different approaches are required to estimate the IMF in different mass regimes. The high-mass end of the IMF is primarily constrained in young stellar populations found in late-type galaxies (LTGs), where the presence of massive stars and gas creates strong nebular emission lines \citep{kennicutt_1998}. Conversely, the low-mass end is generally constrained in old stellar populations within ETGs, which can be approximated as a simple stellar population (SSP, e.g., \citealt{navarro_2019}). These populations, being predominantly old, are composed mostly of low-mass stars, rendering the contribution of high-mass stars negligible.

In composite stellar populations, where old and young stars coexist, studying the low-mass end of the IMF is particularly challenging. First, light from low-mass stars is outshined by massive stars \citep[e.g.,][]{BruzualCharlot_2013,conroy_2012}, diluting the spectral signatures of the low-mass IMF. Second, in young stellar populations, IMF-sensitive absorption features are subject to degeneracies with SFH and chemical enrichment, as changes in age and metallicity can mimic IMF variations. In contrast, in old simple stellar populations, long-lived low-mass stars dominate the light, reducing such degeneracies and making IMF constraints more straightforward. As a result, there are few observational studies estimating the full IMF using stellar population synthesis (SPS) methods beyond the Milky Way. A notable example is \cite{denbrok_2024}, who estimate the IMF shape for 25 cluster ETGs by combining methods: constraining the low-mass end via SPS and the high-mass end through chemical evolution modelling \citep[][]{Dahlgren2018}. By fitting observed chemical abundances, they recover the high-mass IMF slope under the assumption that massive stars leave a chemical imprint on stellar populations through the production of $\alpha$-elements (e.g., C, O, Ne, Mg, Si).

These advances now open the way for a large-scale extragalactic study capable of simultaneously constraining both the low- and high-mass ends of the IMF across a statistically significant sample of star-forming galaxies. Addressing this gap is the central goal of this work. \citet[hereafter MN24]{navarro_2024} introduced a method that accounts for an extended SFH when estimating the low-mass end of the IMF in young stellar populations, thereby reducing the degeneracies introduced by multi-age stellar populations in LTGs. In this work, we build upon the MN24 methodology to constrain the low-mass end of the IMF, while simultaneously estimating the high-mass end using the logarithm of the H$\alpha$ equivalent width (\EWha) versus $g-r$ colour parameter space, referred to as the Kennicutt diagnostic \citep[][]{kennicutt_1983,kennicutt_1994,kennicutt_1998}, following the methodology of \cite{gunawardhana_2011} and \cite{Salvador_2025} applied to star-forming galaxies from the new Hector Galaxy Survey \citep{2024SPIE13096E..0DB}. Combining these two complementary approaches, we aim to perform the first systematic, large-scale, simultaneous study of both the low- and high-mass ends of the IMF in relatively nearby star-forming galaxies ($z \lesssim 0.1$).

The structure of this paper is as follows. In \S \ref{sec:data}, we present the Hector data products and describe the sample selection. In \S \ref{sec:method}, we outline both the MN24 method for estimating the low-mass end IMF slope and the Kennicutt-based method for estimating the high-mass end. Our results are presented in \S \ref{sec:results}. In \S \ref{sec:discussion}, we discuss the limitations of our approach, compare our results with the literature and explore their implications. Finally, in \S \ref{sec:conclusion}, we summarise our key findings and conclusions. Throughout we assume cosmological parameters of $H_0=70\,$km\,s$^{-1}$\,Mpc$^{-1}$, $\Omega_M=0.3$, $\Omega_\Lambda=0.7$ and $\Omega_{\rm{k}} = 0$.

\section{DATA AND SAMPLE}
\label{sec:data}


\subsection{The Hector Galaxy Survey}
\label{sec:data_Hector}

Our analysis is built on observations from the Hector Galaxy Survey \citep[][]{2024SPIE13096E..0DB, Oh_2025}, the latest and ongoing integral field spectroscopy (IFS) project on the 4-metre Anglo-Australian Telescope (AAT). The Hector survey is designed as the major successor to the Sydney-AAO Multi-object Integral-field spectrograph (SAMI) Galaxy Survey \citep{croom_2012,bryant_2015} targeting up to 15,000 galaxies at redshifts $z < 0.1$ to enable statistical studies of galaxy evolution through resolved kinematics, stellar populations, and star formation.

Hector employs 21 fused fibre bundles (hexabundles; \citealt{Bland_Hawthorn_2011,Bryant_2014,2018SPIE10706E..63B,2019SPIE11115E..09W, 2020SPIE11447E..8GW, 2023MNRAS.522.4310W}) deployed across a 2-degree field of view, each with a high ($\sim$75\%) fill factor. The galaxy bundles contain between 61–169 fibres of 1.6 arcsec diameter, corresponding to 14–26 arcseconds in total IFU diameter, and two additional 37-core bundles are dedicated to secondary standard stars for simultaneous calibration. Approximately 70\% of survey galaxies are observed out to at least 2 effective radii ($R_e$). 

These fibre bundles feed two spectrographs: the established AAOmega instrument \citep{Sharp_2006} and the newly commissioned, dedicated Hector spectrograph called Spector \citep{10.1117/12.2314436,10.1117/12.2629649,2024SPIE13096E..0DB}. In total, 855 fibres are fed to Spector and 819 to AAOmega. This dual-instrument setup enables a combination of spatial coverage and spectral resolution. Spector is a fixed-format dual-arm instrument covering 3750–7800 \AA\ (3750–5850 \AA\ in the blue and 5750–7800 \AA\ in the red (\citealt{Oh_2025}; Tuntipong et al., in preparation). The median FWHM of the instrumental line spread function is 1.40$^{+0.09}_{-0.08}$ \AA\ in the blue and 1.20$^{+0.03}_{-0.04}$ \AA\ in the red, corresponding to $\sigma \sim 36.9$ km s$^{-1}$ (blue) and $\sigma \sim 22.5$ km s$^{-1}$ (red). The nominal spectral resolution elements are 0.13 nm in both arms, corresponding to $R\sim 3429$ in the blue and $R\sim 5667$ in the red. This combination of spectral resolution and throughput is critical for robust modelling of stellar populations and IMF-sensitive absorption features.

The Hector data were reduced using the 2dFdr pipeline, following an updated version of the SAMI Galaxy Survey reduction strategy. The process includes bias and overscan correction, bad pixel masking and cosmic ray rejection, fibre tracing and extraction, scattered-light subtraction, flat-fielding, fibre-to-fibre throughput correction, and sky subtraction. Wavelength calibration is performed using a new two-dimensional fitting method that models the solution across the full detector, providing improved accuracy and stability over traditional per-fibre approaches. A full description of the data reduction procedures is presented in \citealt{Oh_2025}.

All data used in this analysis were obtained with the Spector spectrograph only, given its higher spectral resolution and throughput compared to AAOmega, making it best suited for modelling stellar populations and IMF-sensitive absorption features.

\subsection{Sample selection}
\label{sec:data_sample}

To constrain the full IMF shape, we require spectra that satisfy the selection criteria of both estimation methods: galaxies with ongoing star formation with a strong H$\alpha$ emission (\lEWha $\gtrsim 0.8$ \AA) to probe the high-mass end, and a sufficiently high signal-to-noise ratio (S/N) stellar continuum to robustly measure IMF-sensitive absorption features at the low-mass end.

Following the spectroscopic classification scheme of \citet{owers_2019}, galaxies are classified as star-forming using the emission-line diagnostics introduced by \citet{Baldwin_1981} (BPT; Baldwin, Phillips \& Terlevich) diagrams, specifically the [NII]/H$\alpha$ vs [OIII]/H$\beta$ plane with the demarcation lines from \citet{kewley_2006} and \citet{kauffmann_2003}. In this classification scheme, a galaxy is considered star-forming if at least 10$\%$ of its spaxels (with S/N(4100\,\AA) $>$ 3\,pix$^{-1}$) are classified as either star-forming or intermediate (those spaxels that have line ratios that fall between the star-forming and non–star-forming diagnostic boundaries). To ensure sufficient spatial coverage and signal to reliably characterise the IMF across the galaxy, we further restrict our sample to galaxies with at least 50 spaxels classified as star-forming.

\begin{figure*}
\centering
\includegraphics[width=0.9\linewidth]{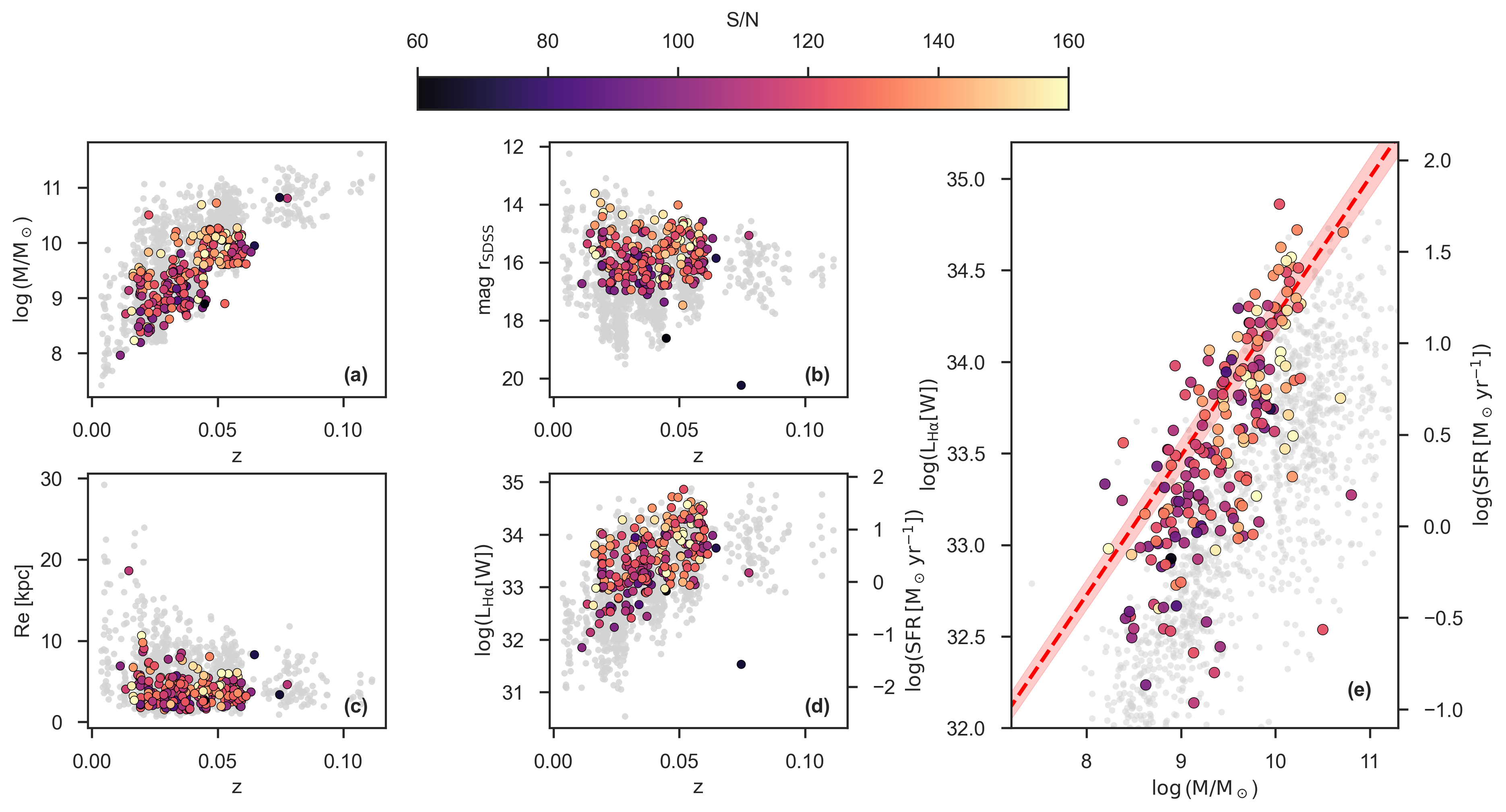}
\caption{Panel a) shows the stellar mass distribution of the sample; panel b) displays the SDSS \textit{r}-band magnitudes; panel c) presents the effective radii; and panel d) shows the logarithmic H$\alpha$ luminosity (in Watts) as a function of redshift ($z$). Panel e) illustrates the relation between H$\alpha$ luminosity (and SFR) and stellar mass, with points colour-coded by the median spectral S/N. The S/N was computed as the median ratio of the observed spectral flux (smoothed to the MILES resolution) to the standard deviation of the residuals, where the residuals are defined as the difference between the observed spectrum and the \texttt{pPXF} fit, measured over two continuum regions: 6066.6–6141.6 $\AA$ and 6422–6455 $\AA$. The red dashed line in panel e) represents the star-forming main sequence (SFMS) from \citet{Renzini_2015}, converted to H$\alpha$ luminosity. Grey points show a control sample of $\sim$1300 star-forming galaxies from the SAMI survey used in \citet{Salvador_2025} for comparison. The sample follows the expected SFMS trend across the observed redshift range. Note that this is a subset of the full Hector survey and does not represent the complete galaxy sample.}
\label{fig_data}
\end{figure*}

This criterion yields a sample of 282 local ($0.01 < z < 0.07$) star-forming galaxies, spanning a stellar mass range of $8.2 < \log(M/M_\odot) < 10.8$ and star formation rates (SFR) of $-2.57 < \log(\text{SFR}~[\text{M}_\odot/\text{yr}]) < 0.11$ (assuming a Salpeter IMF). Figure~\ref{fig_data} provides a visual summary of the physical properties of our sample. As shown in panel e), our galaxies populate the star-forming main sequence (SFMS), while panel a) illustrates how galaxies are distributed in stellar mass and redshift, with low-mass systems primarily found at lower redshifts. Overall, the sample provides a broad coverage of the physical parameter space of local star-forming galaxies.

The final working sample comprises 214 galaxies, after excluding 68 systems outside the validity range of the high-mass IMF method, as discussed in Section~\ref{sec:method_highmass}.

\subsection{Stacking}
\label{sec:data_stacking}

Because IMF-driven variations in absorption features can be very subtle (see Figure~\ref{fig_MCMC_IMF_spec}), robust detection requires a high S/N spectrum, ideally $\mathrm{S/N} \gtrsim 100$ per pixel, though reliable constraints can still be obtained for $\mathrm{S/N} \gtrsim 80$ (MN24). To achieve an optimal S/N, we stack all star-forming spaxel spectra within each galaxy using an inverse-variance weighted sum, following \citet{Allen_2015} and \citet{sharp_2015} for SAMI data,
\begin{equation}
F_\lambda = \sum_i w_{\lambda,i} \, f_{\lambda,i},
\end{equation}
where $f_{\lambda,i}$ is the flux of spaxel $i$ at wavelength $\lambda $, and $ w_{\lambda,i}$ corresponds to the weight cube described in \cite{sharp_2015}, i.e., the inverse of the variance of each spectral pixel. These weights effectively give higher weight to spaxels with higher signal-to-noise.

Before stacking, we identify and replace outlier pixels that could bias our measurements, mainly skyline residuals affecting the pseudo-continuum (blue and red sidebands) and the central spectral-feature bands listed in Table~\ref{table_lines}. To this end, we apply a $3\sigma$ clipping, ensuring that only outlier pixels are replaced, while the absorption signatures of the central bands are preserved. The replaced pixel fractions across the full sample vary across indices, peaking for Fe5335 (median 15.6\%) and Mgb5177 (median 8.27\%). Therefore, metallicity estimates based on the most affected indices should be interpreted with caution\footnote{We repeated the analysis by masking pixels rather than replacing, finding negligible offsets and no significant impact on the inferred IMF slopes or metallicities.}. Outlier pixels are replaced with stellar model fits (Quattropani, in prep.) obtained using \texttt{pPXF} \citep{Cappellari_2004,Cappellari_2017,Cappellari_2023} and X-Shooter templates \citep{Chen_2014}, following \cite{owers_2019}.

Properly propagating the covariance is essential to avoid underestimating the uncertainties in the final stacked spectrum, as neighbouring spaxels are not statistically independent due to the resampling process in the data reduction \citep[e.g.,][]{husemann_2013,garciabenito_2015,gunawardhana_2020}. Since our analysis relies on fitting spectral absorption features using SPS models, it is crucial that the errors in the stacked spectrum are accurately estimated; otherwise, the derived stellar population parameters may be significantly over- or underestimated. The associated variance was computed by reconstructing the full covariance matrix from the reduced covariance data stored in the Hector data cubes. For each wavelength slice, the total variance was estimated as
\begin{equation}
\sigma_\lambda^2 = \sum_{i,j} w_{\lambda,i} \, w_{\lambda,j} \, \mathrm{Cov}_{\lambda}(i,j),
\end{equation}
where the sum runs over all spaxel pairs within a $5 \times 5$ region centred on each star-forming spaxel, and $\mathrm{Cov}_\lambda(i,j)$ denotes the spatial covariance between spaxels $i$ and $j$ at wavelength $\lambda$. 

Finally, the stacked blue and red spectra were interpolated onto a common wavelength grid, using the finer sampling of the red arm. This ensures that absorption lines falling in the overlapping wavelength region are properly aligned. The fluxes in overlapping regions were then averaged. The resulting stacked spectrum reaches an average S/N of $\sim$ 119 with a standard deviation of 21.

\begin{table*}[h!]
\centering
\caption{Spectral regions used to estimate the low-mass end slope of the IMF, as defined in the Lick/IDS system \citep{worthey_1994}. For each absorption feature, the table lists the central bandpass used to measure the index strength, together with the adjacent blue and red pseudo-continuum sidebands. These sidebands define the local continuum around each feature, ensuring accurate and consistent index measurements. All wavelengths are given in Angstroms.}
\begin{tabular}{lccc}
\hline
\textbf{Name} & \textbf{Blue [\AA]} & \textbf{Line [\AA]} & \textbf{Red [\AA]} \\
\hline
Mgb5177  & 5142.625 -- 5161.375 & 5160.125 -- 5192.625 & 5191.375 -- 5206.375 \\
Fe5270   & 5233.150 -- 5248.150 & 5245.650 -- 5285.650 & 5285.650 -- 5318.150 \\
Fe5335   & 5304.625 -- 5315.875 & 5312.125 -- 5352.125 & 5353.375 -- 5363.375 \\
TiO$_1$     & 5816.825 -- 5849.125 & 5936.625 -- 5994.125 & 6038.625 -- 6103.625 \\
TiO$_2$     & 6066.625 -- 6141.625 & 6189.625 -- 6272.125 & 6372.625 -- 6415.125 \\
\hline
\end{tabular}
\label{table_lines}
\end{table*}

\subsection{Deriving H$\alpha$ luminosities and equivalent widths}
\label{sec:data_HaLum}

We obtain the extinction-corrected H$\alpha$ luminosities ($L(\mathrm{H}\alpha)$) for each star-forming spaxel using \texttt{spaxelsleuth} \citep[][]{Zovaro_2024}, a Python package for analysing IFU data. Emission-line fluxes are corrected for dust attenuation on a spaxel-by-spaxel basis and subsequently converted into luminosities by multiplying by $4\pi D_L^2$, where $D_L$ is the luminosity distance.

Dust attenuation is estimated from the Balmer decrement using the observed $\mathrm{H}\alpha/\mathrm{H}\beta$ ratio and assuming an intrinsic value of $(\mathrm{H}\alpha/\mathrm{H}\beta)_0 = 2.86$. We adopt the \citet{Fitzpatrick_2007} reddening law and compute the colour excess as $E(B{-}V) = [2.5/(k(\mathrm{H}\beta)-k(\mathrm{H}\alpha))]\log_{10}[(\mathrm{H}\alpha/\mathrm{H}\beta)/2.86]$, from which the V-band extinction is obtained as $A_V = R_V\,E(B{-}V)$ with $R_V=3.1$. The extinction correction is applied on a spaxel-by-spaxel basis to all emission lines as $F_{\mathrm{corr}} = F_{\mathrm{obs}}\,10^{0.4 A_\lambda}$, and flux uncertainties are propagated accordingly.

The global luminosity of a galaxy is then calculated by summing the luminosities of all its star-forming spaxels. To derive the H$\alpha$ luminosity surface density, $\Sigma_{L(\mathrm{H}\alpha)}$, we divide each spaxel luminosity by its projected area, i.e., the bin size in units of kpc$^2$.

The equivalent width of H$\alpha$, \EWha, is measured from the stacked spectra as the ratio between the H$\alpha$ emission-line luminosity and the local pseudo-continuum, defined as the mean flux density in the rest-frame range 6500–6540~\AA. Both the emission and continuum luminosities are estimated with \texttt{spaxelsleuth}.

\section{METHODS}
\label{sec:method}


While the IMF has been extensively studied in the Milky Way (see \citealt{bastian_2010} and references therein), direct constraints outside our Galaxy remain limited. Estimating both the low- and high-mass ends of the IMF provides a more complete picture of the stellar mass distribution, offering key insights into how the star formation process varies across environments. This section describes two approaches to estimate the IMF slopes: the SPS method from MN24 for the low-mass end IMF slope ($\alpha_{\rm low}$), and the \lEWha\ vs $g-r$ diagnostic of \citep{kennicutt_1983} for the high-mass end IMF slope ($\alpha_{\rm high}$).

\subsection{Estimating the low-mass end IMF slope}
\label{sec:method_lowmass}


An SPS-based IMF estimation method exploits the sensitivity of certain absorption features to stellar surface gravity, which is closely linked to stellar mass. Since the IMF governs the stellar mass distribution, these features provide indirect constraints on the low-mass IMF slope. SPS relies on SSP models that predict integrated spectra for given age, metallicity, abundance ratios, and IMF. By finding the combination of SSP templates that best reproduces the observed IMF-sensitive features, this approach constrains both the stellar population properties and the low-mass IMF slope.

Although we refer to these absorption features as ``IMF-sensitive,’’ they are in fact influenced by multiple factors, primarily stellar surface gravity and effective temperature, which along an isochrone translate into a dependence on stellar mass (e.g., MN24). Consequently, the strength of a given feature reflects the combined effects of the IMF, stellar age, metallicity, and abundance ratios. Throughout this paper, we use the term ``IMF-sensitive’’ in this practical sense.

\subsubsection{Stellar population synthesis models}
\label{sec:SPSm}

The SSP models adopted in this work are based on the MILES models \citep{Vazdekis_2010}, which incorporate variations in $\alpha$-element abundances \citep{Vazdekis_2015}. These models are built from BaSTI isochrones \citep{Pietrinferni_2004,Pietrinferni_2006}, the empirical MILES stellar library \citep{Sanchez-Blazquez_2007}, and synthetic spectra from \citet{Coelho_2005}, delivering intermediate-resolution spectra (2.51~\AA) over the $\sim$3500–7400~\AA \, wavelength range.

The subset of MILES models with enhanced [$\alpha$/Fe] spans a stellar metallicity range of [M/H] = –2.27 to +0.40. These models include two $\alpha$-element abundance patterns: [$\alpha$/Fe] = 0.0 and +0.4. We refer to [$\alpha$/Fe] as [Mg/Fe], since magnesium is the only $\alpha$-element directly constrained by our indices. 

The MILES models do not self-consistently include variations in individual elemental abundances. Following MN24, we therefore introduce Ti abundance variations in a differential way by applying the response functions of \citet{Conroy_2012b} to the relevant spectral features. In this framework, [Ti/Fe] is included only to capture residual sensitivity of the indices to Ti, rather than to provide a robust abundance determination. Accordingly, [Ti/Fe] is treated as a nuisance parameter, and its inferred values are not interpreted as precise measurements in this work.

Within this modelling framework, the IMF is parametrised as a single power law with logarithmic slope $\Gamma$ ranging from 0.3 to 3.3 ($-4.3<\alpha_{\rm low}<-1.3$). As discussed in MN24, the lack of observational constraints on the low-mass end of the IMF in young stellar populations motivates this parametrisation choice, since a unimodal parametrisation provides both a practical and physically consistent simplification.

The standard MILES models are limited to ages above 30~Myr. However, in star-forming galaxies, much younger stellar populations can significantly contribute to the integrated light. To address this, MN24 complement the $\alpha$-variable MILES grid with the super-young MILES extension from \citet{Asad_2017}, which covers ages between 6.3~Myr and 63~Myr using \citet{Bertelli_1994} isochrones. To merge both model sets into a regular grid, the super-young models are linearly interpolated to match the metallicity sampling of the $\alpha$-variable models ($\sim$ 0.4–0.6 dex). Since the impact of abundance variations decreases with decreasing stellar age \citep[see e.g.,][]{Vazdekis_2015}, the abundance sensitivity is extended to the youngest ages by applying metallicity-dependent response functions. These functions are computed as the ratio between [$\alpha$/Fe] = +0.0 and [$\alpha$/Fe] = +0.4 models at fixed metallicity for the youngest $\alpha$-variable MILES spectra, and then applied to the super-young models. Below [M/H] = –0.4, where super-young models have [$\alpha$/Fe] $\approx$ 0.4, the correction produces predictions at [$\alpha$/Fe] $\approx$ 0.0; above this metallicity, the correction is reversed to reach [$\alpha$/Fe] $\approx$ 0.4.

The final combined stellar population model grid covers ages from 6.3 Myr to 14 Gyr, spans a wide range in metallicity, and includes variations in [M/H], [Mg/Fe], [Ti/Fe], and the IMF slope.

\subsubsection{Fitting for stellar population parameters}
\label{sec:MCMC}

The IMF slope estimation method introduced in MN24 proceeds in two main steps: first deriving the SFH, and then fitting the stellar population parameters. In the first step, we use \texttt{pPXF} to determine the best-fitting combination of multi-age stellar populations. When estimating the SFH, we restrict the template set to SSP models varying only in age, [M/H], and IMF slope, since constraining the age distribution is the primary goal at this stage.

Prior to fitting, both the observed spectra and the templates are normalised and log-rebinned to ensure a consistent sampling in velocity space. The Hector spectra have an instrumental resolution of $\mathrm{FWHM_{gal}} \sim 1.3$,\AA, while the MILES templates have $\mathrm{FWHM_{MILES}} \sim 2.51$,\AA. To match the two, we convolve the observed spectra with a Gaussian kernel of dispersion as done in \cite{owers_2019}
\[
\sigma_{\rm ker} = \sqrt{\sigma_{\rm MILES}^2 - \sigma_{\rm gal}^2}.
\]
In this work, we assume a constant instrumental resolution across the full wavelength range (1.3\AA), and therefore apply the same kernel at all wavelengths. This assumption is discussed in Section \ref{sec:discussion_limitations}.


We perform an initial \texttt{pPXF} fit on our normalised spectra over the wavelength range $3550$–$7000,\mathrm{\AA}$ using a $3\sigma$-clipping routine to identify outlier pixels, which are masked and excluded in any subsequent fit, following the approach described in \citet{Cappellari_2023}. The NaD line is also masked (5890–5896 \AA), as it is highly sensitive to the interstellar medium (\citealt{Poznanski_2012,Sternberg_2013}). This first fit models the continuum with a multiplicative 12 degree Legendre polynomial, and its purpose is to detect and remove outliers. In the second run, using the cleaned spectrum, the first two moments of the line-of-sight velocity distribution (LOSVD), namely the velocity ($v$) and velocity dispersion ($\sigma$), are fit simultaneously while modelling and subtracting the ionised-gas emission to isolate the stellar continuum. This iterative procedure leads to an improvement in the reduced chi-squared statistic ($\chi^2/\nu$).

A third \texttt{pPXF} run is then carried out with the kinematics fixed from the previous step, applying a linear regularisation parameter of 20 across the age–metallicity–IMF parameter space\footnote{We verified that varying this regularisation within reasonable bounds does not significantly alter the derived $\alpha_{\rm{low}}$.}. This regularisation enforces a smoother distribution of template weights, effectively producing a non-delta-function SFH. This step yields a first-order estimate of the SFH and a luminosity-weighted mean age for each spectrum. To quantify uncertainties in the recovered SFH, we repeat this fit ten times per galaxy, adding Gaussian noise to the spectrum in each iteration with an amplitude equal to the residuals of the best-fit model. The uncertainty in each SFH bin is then estimated as the standard deviation of the ten realisations, providing a first-order measure of the reliability of the recovered SFH.

In the second step, we estimate the stellar population parameters. First we bring the observed spectra and the model templates to a common rest frame and match their total broadening (instrumental plus kinematic) as done in equation 1 in \citet{owers_2019}. The rest-frame correction is applied using $1 + z_{\rm tot} = (1 + z_{\rm gal})(1 + v/c)$, where $z_{\rm gal}$ is the galaxy redshift, $v$ is the velocity relative to $z_{\rm gal}$ obtained in the first step, and $c$ is the speed of light. Each template from the MILES library is then convolved with a Gaussian kernel of wavelength-dependent width:
\begin{equation}
\sigma_\lambda^2 = \left[ \left( \frac{\lambda \sigma_{\rm pPXF}}{\rm c}\right)^2 + \left(\frac{\sigma_{\rm inst}}{1+z} \right)^2 \right] - \sigma_{\rm MILES}^2,
\end{equation}
where $\sigma_{\rm pPXF}$, $\sigma_{\rm inst}$, and $\sigma_{\rm MILES}$ are the velocity dispersion from the first step and the instrumental dispersions of the data and the MILES models, respectively. 

While stellar populations in ETGs are generally more homogeneous in age and metallicity \citep[e.g.,][]{CidFernandes_2005,Gallazzi_2005} and can often be well approximated by a single SSP, this assumption is inadequate for systems with more complex SFHs, such as LTGs. For instance, Figure~2 in MN24 illustrates that a given line-strength index can arise from either a young stellar population with a top-heavy IMF or an old population with a bottom-heavy IMF. This degeneracy can be alleviated by including an age proxy in the SSP fit. The key advantage of the MN24 approach is that it fits the absorption spectra while allowing for an extended SFH, rather than assuming a single-age population. To do this, we build a grid of MILES model spectra, where each template corresponds to a combination of four luminosity-weighted stellar population parameters: [M/H], [Mg/Fe], [Ti/Fe], and the IMF slope. To further refine the fit, the MN24 framework introduces two additional free parameters. The first is a multiplicative factor that scales the contribution of a 10 Myr-old stellar population, designed to capture the flux from luminous supergiant stars ($f_{\rm{RSG}}$). These stars can influence the TiO absorption features, and if unaccounted for, their strong continuum contribution could make the low-mass IMF appear more bottom-heavy than it actually is. By treating this population as an independent burst, the factor provides flexibility to model the relative impact of supergiants on the spectral indices, without implying a physical fraction of the total light. The second parameter is a rescaling factor on the error spectrum ($\Delta_{\rm{err}}$), which allows the model to absorb residual systematics in the data.


The templates that define our model grid are not pure SSP spectra. Instead, they are age-composite populations constructed by combining multiple stellar components of different ages, where the relative weight of each age component is set by the SFH derived in the previous \texttt{pPXF} step. In this sense, each model represents an extended SFH rather than a single-age population.

All age components within a given template share the same values of the other stellar population parameters ([M/H], [Mg/Fe], [Ti/Fe], and IMF slope). This approximation neglects any temporal evolution in abundance ratios or metallicity across the SFH, but it is a necessary and reasonable simplification, as introducing fully time-dependent chemical enrichment would substantially increase the model complexity. Nevertheless, constraining the age distribution remains the dominant factor in mitigating IMF–age degeneracies, since age variations drive the largest changes in the absorption-line spectra. This limitation is discussed further in Section~\ref{sec:discussion_limitations_lowmass}.

We apply a full index fitting (FIF) technique based on a Bayesian Markov Chain Monte Carlo (MCMC) algorithm, implemented using the \texttt{emcee} sampler \citep{Foreman-Mackey_2013}. This method compares the SFH-weighted model spectra with the observed data by maximising the following likelihood function:
\begin{equation}
\ln(\mathbf{O}|\mathbf{M})=-\frac{1}{2}\sum_n\left[\frac{(O_n-M_n)^2}{\sigma^2_n}-\ln\left(\frac{1}{\sigma^2_n}\right)\right],
\end{equation}
where $O_n$ and $M_n$ are the observed and model fluxes at the $n$th pixel, and $\sigma_n$ is the uncertainty in the observed flux. Unlike traditional full spectral fitting, FIF focuses on specific regions of the spectrum where the diagnostic power is strongest and best understood. In our analysis, the fitting is restricted to five selected absorption features: Mgb, Fe5270, Fe5335, TiO$_1$, and TiO$_2$. The wavelength intervals around each feature, including their sidebands, are summarised in Table \ref{table_lines}. The sidebands are used to estimate the local continuum level by fitting a low-order polynomial to the flux ratio between the observed and model spectra within these regions. This continuum correction ensures that differences between data and models are not driven by large-scale flux calibration uncertainties or residual mismatches in the continuum shape, but rather by the intrinsic strength of the absorption features themselves.

To illustrate how the models respond to variations in the IMF slope, Figure~\ref{fig_MCMC_IMF_spec} shows model spectra with all stellar population parameters held fixed except for the IMF slope. Although the spectral differences appear modest as the low-mass end flattens, in practice the MCMC simultaneously explores variations in the IMF, SFH, and chemical composition. This amplifies the spectral differences beyond what is visible in the figure and allows for a robust, informed, non-degenerate inference of the IMF.

Using FIF instead of index–index diagram methods offers several advantages. First, by limiting the analysis to specific indices, we reduce sensitivity to continuum shape and flux calibration uncertainties. Second, FIF increases the number of independent observables, as each pixel within an index region contributes separately, providing tighter constraints than traditional index–index approaches. Third, the differential sensitivity of individual pixels within a feature to age, metallicity, and IMF slope helps to disentangle degeneracies between these parameters. Compared to full spectral fitting, FIF focuses only on selected spectral features, which can reduce the impact of mismatches in continuum modelling while still retaining detailed information within the indices.

In the MCMC module, we adopt Gaussian priors on [M/H] and the IMF slope, centred on the values inferred from the initial \texttt{pPXF} fits, with a standard deviation of 0.3. A narrow Gaussian prior ($\sigma = 0.01$) is applied to [Ti/Fe], fixed at solar abundance ([Ti/Fe] = 0.0), to prevent it from reaching the edges of the model grid. This choice has a negligible effect on the inferred IMF slope because TiO features, while mildly sensitive to [Ti/Fe], respond primarily to surface gravity and thus to the IMF itself. Fixing [Ti/Fe] therefore helps stabilise the fit without biasing the IMF or other stellar population parameters (see Appendix~\ref{sec:appendix_TiFe}).

The fitting procedure is repeated independently for each of the ten \texttt{pPXF} realisations. Figure~\ref{fig_MCMC} shows an example of the resulting posterior distribution. Final parameter estimates and their uncertainties are derived from the combined posterior samples of all MCMC realisations. For each parameter, we report the median value (50th percentile) as the best-fitting estimate, while the 16th and 84th percentiles define the lower and upper uncertainties, respectively (see Figure \ref{fig_posterior}). In other words, the parameter estimate is taken as the median, and the uncertainty reflects the range encompassing 68\% of the posterior distribution around this median. Across the ten MCMC realisations, the fits achieve an average reduced $\chi^2$ of $0.68$ with a standard deviation of $0.44$, indicating statistically acceptable fits. We refer the reader to \cite{navarro_2019}, \cite{navarro_2021}, and MN24 for a full and detailed description of the method.

Notice that $\alpha_{\rm{low}}$ mainly probes the population of M-dwarf stars, which have masses in the range $\sim 0.08$–$0.6 \, M_\odot$ \citep{Henry_2024}. While the optical continuum is dominated by massive stars, the TiO absorption features used in this analysis arise exclusively in the atmospheres of low-mass stars (M dwarfs), meaning that the inferred slope effectively probes this low-mass stellar population. Because these stars are long-lived, the inferred slope reflects a luminosity-weighted average over several Gyr of star formation, rather than the properties of a single-age population. From the SFHs recovered with \texttt{pPXF}, the typical luminosity-weighted age of the populations contributing to the optical absorption lines is $\sim 3.5$ Gyr ($\sigma= 1.7~\rm Gyr$), indicating that the low-mass slope traces a time-averaged contribution from intermediate-age to old stellar populations.

To assess the robustness of the method, we apply it to 94 quiescent galaxies from the Hector survey (see Section~\ref{sec:appendix_QS}) and find results that are broadly consistent with previously reported IMF–velocity dispersion trends \citep[e.g.][]{Labarbera_2013, ferreras_2012}.

\subsubsection{The spectral features}
\label{sec:specfeat}

\begin{figure*}
\centering
\includegraphics[width=0.8\linewidth]{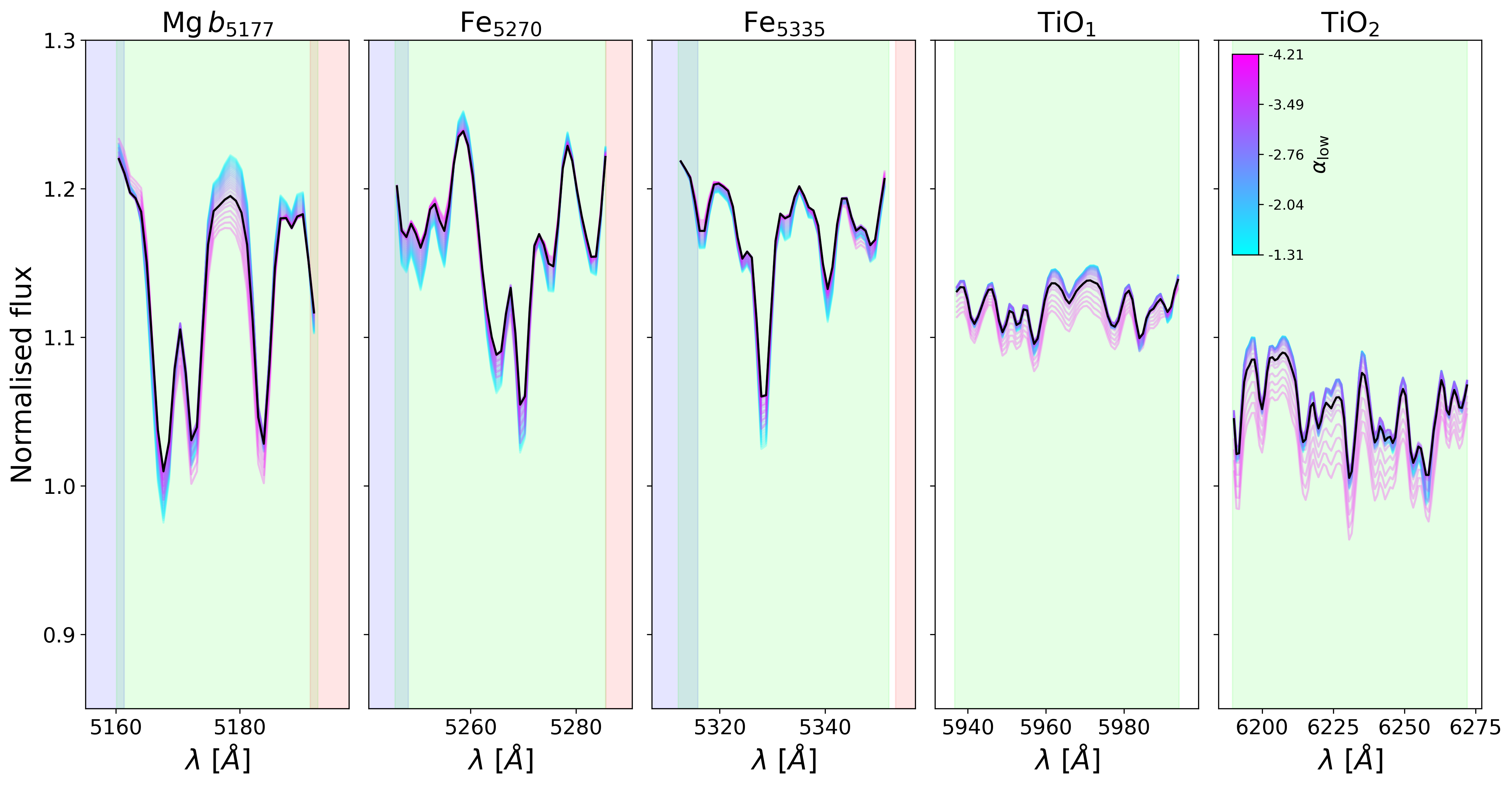}
\caption{Illustration of the constraining power of IMF-sensitive spectral features (from left to right: Mgb5177, Fe5270, Fe5335, TiO$_1$, and TiO$_2$) for galaxy W17670457407415. The black curve shows the best-fitting model spectrum obtained from the MCMC module. The coloured curves represent model spectra where all parameters are fixed to their best-fitting values, except for the IMF slope, which is varied. The colour scale indicates the IMF slope, with lighter blue corresponding to more bottom-light IMFs and pink to more bottom-heavy ones. Shaded regions mark the wavelength intervals used in the analysis: blue for the blue pseudo-continuum sidebands, green for the central line regions, and red for the red sidebands, as defined in Table \ref{table_lines}.}
\label{fig_MCMC_IMF_spec}
\end{figure*}

\begin{figure*}
\centering
\includegraphics[width=0.8\linewidth]{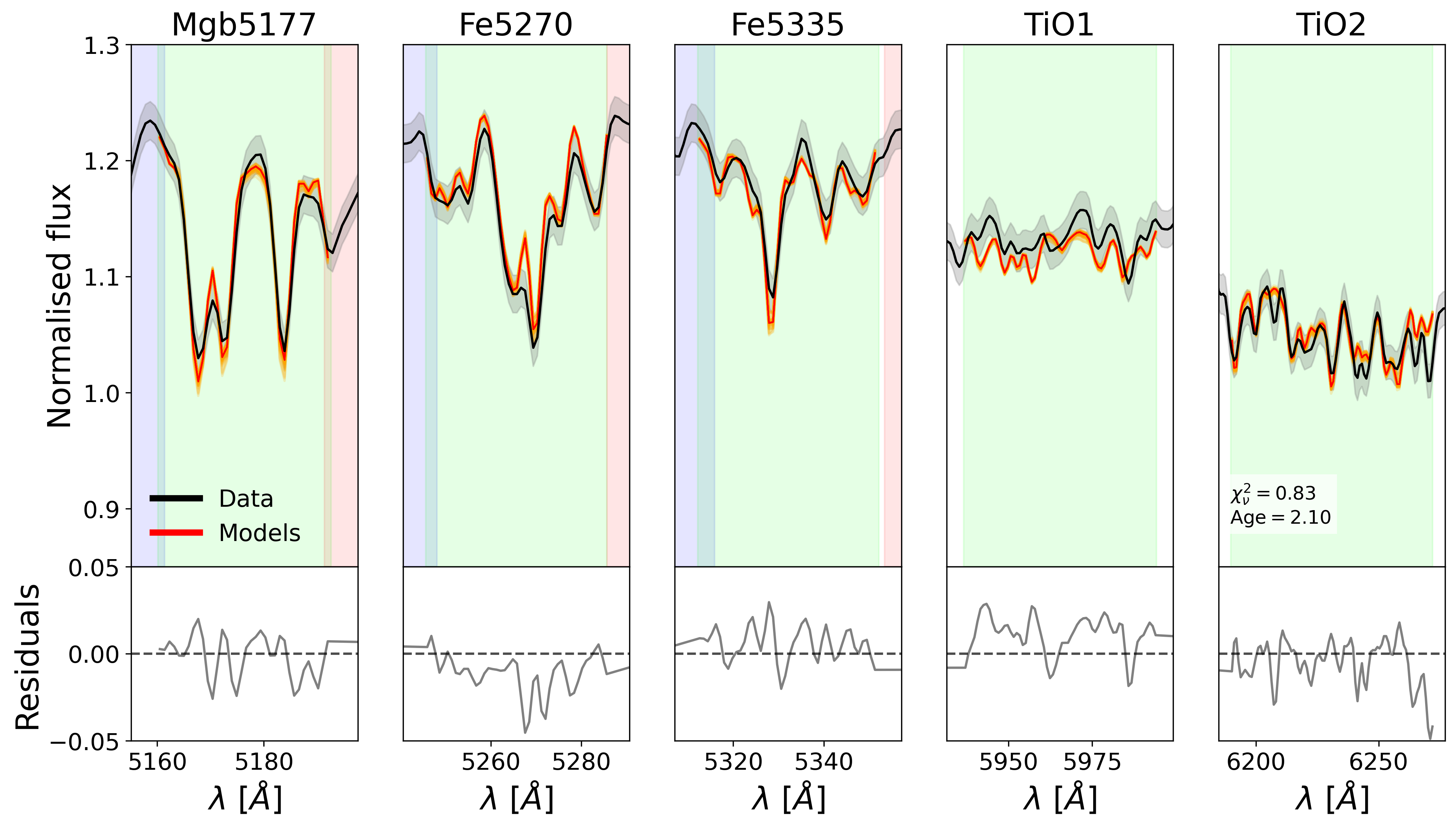}
\caption{Example of five IMF-sensitive absorption features in the spectrum of galaxy W17670457407415 (black line). The grey shaded regions show the $\pm$1$\sigma$ observational uncertainty of the spectrum. The red line indicates the best-fitting model (median of the posterior MCMC samples), while the orange lines correspond to 50 randomly drawn MCMC realisations (after the burn-in) illustrating model variance. The shaded blue, green, and red regions mark the blue pseudo-continuum sidebands, central line regions, and red sidebands, respectively, as defined in Table \ref{table_lines}. The fit simultaneously constrains the luminosity-weighted stellar population parameters ([M/H], [Mg/Fe], [Ti/Fe] and IMF slope) by maximising the Gaussian likelihood between the observed and model flux within each feature’s bandpass. Residuals (bottom panels) are shown for each line, remaining smaller than 5\% of the normalised flux. The rightmost panel reports the luminosity-weighted stellar age and the reduced $\chi^2$ ($\chi^2/\nu$) of the fit.}
\label{fig_MCMC}
\end{figure*}

\begin{figure}
\centering
\includegraphics[width=0.9\linewidth, height=0.7\textheight, keepaspectratio]{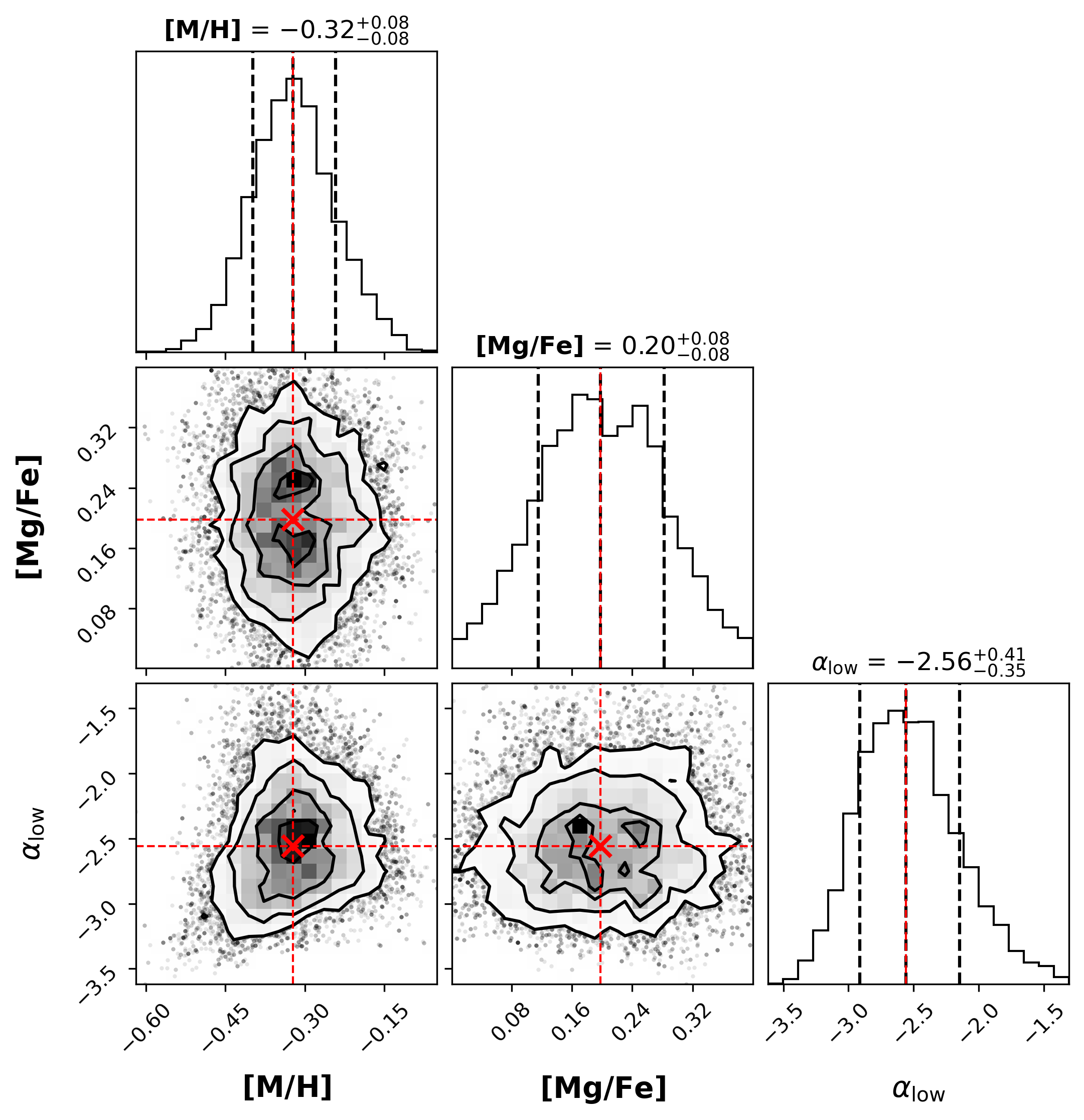}
\caption{Posterior distributions for the stellar population parameters [M/H], [Mg/Fe] and low-mass end IMF slope ($\alpha_{\rm low}$), obtained from the Bayesian full-index fitting method applied to ten SFH realisations per galaxy for galaxy W17670457407415. The vertical dashed lines indicate the 16th to 84th percentile uncertainties derived by combining the posteriors from all realisations. The red cross marks the selected parameter values.}
\label{fig_posterior}
\end{figure}

In our analysis, the IMF-sensitive features are TiO$_1$ and TiO$_2$. As shown in Figure \ref{fig_MCMC_IMF_spec}, both exhibit a strong response to IMF variations, making them particularly powerful for constraining the low-mass end of the IMF. The Mgb and iron indices (Fe5270 and Fe5335) are not themselves direct IMF tracers, but are essential for constraining metallicity and [$\alpha$/Fe], thereby alleviating degeneracies that would otherwise bias IMF estimates (e.g., \citealt{worthey_1994,thomas_2003,spiniello_2014}). 

Several studies have demonstrated that titanium oxide features, particularly TiO$_1$ and TiO$_2$, provide strong leverage on the low-mass end of the IMF, as they arise in the cool atmospheres of low-mass stars (mainly M dwarfs; \citealt{Allard_1997}). Consequently, in the wavelength regions of these TiO absorption bands, the flux originates almost entirely from stars at the low-mass end of the IMF. \citet{spiniello_2012} and \citet{spiniello_2014} showed that TiO$_2$ is especially effective at decoupling IMF effects from age, metallicity, and abundance ratios, while also correlating with NaI in massive ETGs, thus supporting its interpretation as an IMF-sensitive indicator rather than interstellar contamination. Similarly, \citet{navarro_2015b} emphasised that TiO$_2$ is a direct tracer of the dwarf-to-giant ratio, with clear radial gradients consistent with IMF variations.

While the MN24 framework incorporates the NaD doublet as another IMF indicator, we exclude it and rely instead on TiO molecular features, as NaD is strongly affected by interstellar absorption (e.g., \citealt{jeong_2013}). In our sample, we find that the effect of adding the NaD doublet to the MCMC framework, and therefore [Na/Fe], results in more bottom-heavy IMF slopes (see Appendix~\ref{sec:appendix_NaD}).

Several other IMF-sensitive features have been employed in the literature. One of the most widely discussed is the Wing–Ford band (e.g., \citealt{conroy_2012,labarbera_2016}). However, this index lies in a spectral region strongly affected by telluric absorption and sky residuals, limiting its practical use in large galaxy surveys. Compared to other commonly adopted IMF-sensitive indicators, such as the NaI doublet or the CaII triplet, TiO indices are less prone to contamination from interstellar absorption and, to some extent, from abundance ratio variations, although residual degeneracies with [$\alpha$/Fe] and stellar effective temperature remain (e.g., \citealt{conroy_2012,spiniello_2012,Labarbera_2013}). As shown in \ref{sec:appendix_NaD}, using TiO$_1$ and TiO$_2$ individually tends to yield slightly more bottom-heavy IMF estimates for some galaxies, but both remain reasonably strong IMF indicators for most of the sample (e.g., 53.1\% of galaxies show a difference of less than 0.5 in $\alpha_{\rm low}$ when TiO$_1$ is removed, and 59.8\% when TiO$_2$ is removed). Overall, combining both features provides more robust IMF constraints, making TiO$_1$ and TiO$_2$ valuable IMF tracers within the optical spectral range.

\subsection{Estimating the high-mass end IMF slope}
\label{sec:method_highmass}


To constrain the high-mass-end IMF slope, $\alpha_{\rm high}$, we employ the Kennicutt diagnostic \citep{kennicutt_1983,kennicutt_1994}, which uses H$\alpha$ emission as a tracer of recent massive star formation. This method is based on the relation between the logarithm of the H$\alpha$ equivalent width (\lEWha) and the $g-r$ colour: \EWha traces recent star formation through ionisation by massive, short-lived stars, while $g-r$ reflects older, low-mass-dominated populations. Consequently, a galaxy’s location in the \EWha–$g-r$ plane is sensitive to the high-mass IMF slope.

As discussed before, the \EWha\ is defined as the ratio between the H$\alpha$ line luminosity and the local pseudo–continuum. The H$\alpha$ luminosity is corrected for dust extinction using the Balmer decrement, while the continuum is corrected with the \citet{calzetti_2001} reddening curve. The apparent AB magnitudes in the SDSS $g$ and $r$ bands are then computed by integrating the rest-frame galaxy spectra through the corresponding filter response functions. These colours are corrected for dust extinction following the same attenuation law, with the colour excess $E(B-V){\text{gas}}$ estimated from the Balmer decrement. We adopt the commonly used relation $E(B-V){\rm star} = 0.44,E(B-V){\rm gas}$ to account for the lower attenuation of the stellar continuum relative to nebular emission. We note, however, that this ratio varies significantly between galaxies, up to $\sim 1$ (e.g., \citealt{puglisi_2016, shivaei_2020}). An underestimate of $E(B-V){\rm star}$ would make the stellar continuum appear bluer (by up to $\sim 0.27$ mag in extreme cases), potentially leading to systematically higher estimates of $\alpha{\rm high}$.

To link the observables with the IMF slope, we generate a grid of theoretical models using the SPS code \texttt{PÉGASE}.3 \citep{fioc_2019}. The nebular emission in \texttt{PÉGASE} is computed using photo-ionisation models based on the number of Lyman continuum photons emitted by young stars, assuming a spherical, radiation-bounded HII region with constant hydrogen density and no internal dust.

We adopt a similar in slope to the \citet{kroupa_2001} IMF at the low-mass end, fixed at $\alpha_{\rm low} = -1.3$ for $0.08 < \rm M/M_\odot < 0.5$, and a variable high-mass slope $\alpha_{\rm high}$ for $0.5 < \rm M/M_\odot < 120$, ranging between $-4.0$ and $-1.0$ in steps of 0.01. The upper mass cutoff of the models is $M_{\rm max} = 120~\rm M_\odot$. We assume an exponentially declining SFH with an e-folding time of 1.1 Gyr and a constant metallicity $Z=0.02$. The resulting model grid spans a wide range of \EWha\ and $g-r$ values as a function of the high-mass IMF slope.

For a given combination of input parameters (SFH, metallicity, IMF), \texttt{PÉGASE} evolves a stellar population and outputs broadband colours and emission-line equivalent widths at each time step. This enables the construction of evolutionary tracks in the \lEWha vs $g-r$ space, with each track being associated to a different IMF slope. By comparing measurements of \lEWha and $g-r$ with these tracks, an IMF slope can be inferred. 

Since the model grid is discrete, we use inverse distance weighted interpolation (IDWI; \citealt{shepard_1968}) to assign an IMF slope, $\alpha_{\rm{high}}$, to each pair of \lEWha and $g-r$ colour. For a given observed pair $\bm{x} = (g-r,$\lEWha$)$, the interpolated slope $u(\bm{x})$ is calculated as:
\begin{equation}
    u(\bm{x}) = \frac{\sum_{i=1}^N \omega_i(\bm{x}) u_i}{\sum_{i=1}^N \omega_i(\bm{x})},
\end{equation}
where
\begin{equation}
    \omega_i = \frac{1}{d(\bm{x}, \bm{x}_i)^m}.
\end{equation}
Here, $u_i$ is the IMF slope corresponding to the model grid point $\bm{x}_i$, $d$ is the Euclidean distance in the parameter space, and $m$ is the power parameter, which we set to $m = 3$. A higher $m$ gives more weight to the closest neighbours. If $d(\bm{x}, \bm{x}_i) = 0$, we set $u(\bm{x}) = u_i$. For each observed point, we define a circular region in parameter space and include all evolutionary track points that fall within it when performing the interpolation.

\begin{figure}[hbt!]
\centering
\includegraphics[width=0.9\linewidth, height=0.7\textheight, keepaspectratio]{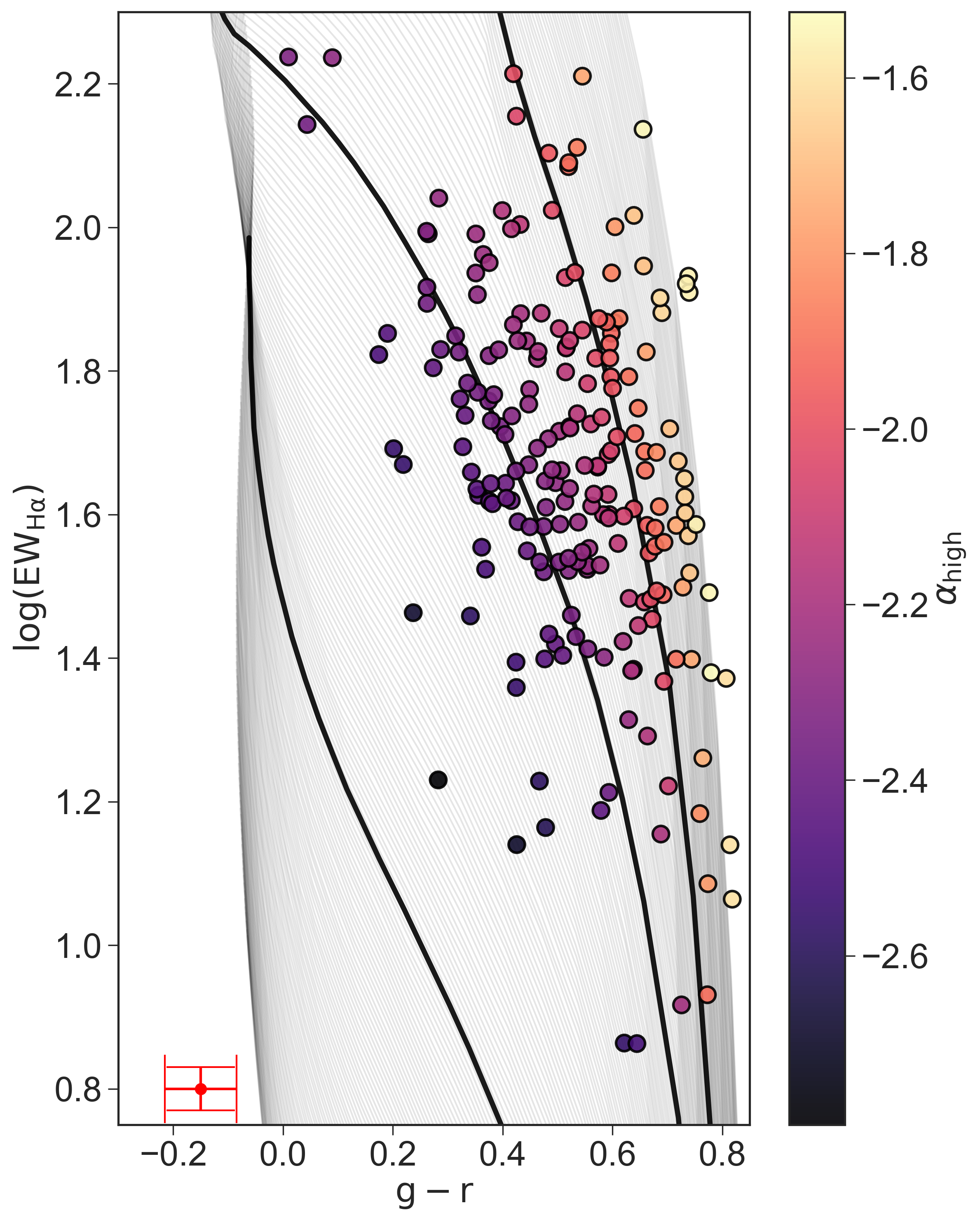}
\caption{Distribution of galaxies in the \lEWha vs $g-r$ diagnostic plane. The black grid shows model predictions from \texttt{PÉGASE}.3, spanning a range of high-mass IMF slopes ($\alpha$ from –1.5 to –4.0). For reference, models with $\alpha_{\rm{high}} = -3$, $-2.35$, and $-2$ (lines from left to right) are highlighted with thicker black lines. Our galaxy sample is overlaid points coloured by $\alpha_{\rm high}$. The red cross in the lower left corner illustrates the median observational uncertainties in colour and \lEWha.}
\label{fig_tracks}
\end{figure}

Figure~\ref{fig_tracks} shows the observed distribution of galaxies in the \lEWha vs $g-r$ plane, overlaid with the model evolutionary tracks. This illustrates how galaxy positions in this diagram are related to different IMF slopes, and highlights the constraining power of this method.

We excluded from our analysis a total of 68 galaxies lying outside the region covered by our evolutionary track models in the \lEWha–$g-r$ plane, as the high-mass-end IMF slope cannot be reliably constrained in this part of the parameter space. The final sample therefore comprises 214 galaxies.

Uncertainties in the inferred high-mass-end IMF slope are estimated by propagating errors in \lEWha and $g-r$. For each object, we define an error box spanning $(g-r \pm \sigma_{g-r})$ and $(\log \mathrm{EW}_{\mathrm{H}\alpha} \pm \sigma_{\log \mathrm{EW}})$. We then generate 100 random points uniformly distributed within this box, estimate the IMF slope for each, and adopt the standard deviation of these slopes as the final uncertainty.

The high-mass-end slope, $\alpha_{\rm{high}}$, is primarily sensitive to stars with masses $\gtrsim 10\, \rm M_\odot$ \citep[][]{kennicutt_1994}, as these are the stars capable of producing the ionising photons traced by the nebular H$\alpha$ emission. Because such massive stars have short lifetimes ($\lesssim 10$ Myr), the inferred slope effectively probes the recent star formation activity over this timescale.

\section{RESULTS}
\label{sec:results}


In this section, we present our low- and high-mass end IMF slopes estimates obtained as described in Sections~\ref{sec:method_lowmass} and \ref{sec:method_highmass}. We first assess whether the low- and high-mass IMF components are correlated, and then examine their relations with key galaxy properties, including stellar mass, H$\alpha$ luminosity, and chemical composition ([M/H] and [Mg/Fe]).

We quantify correlations between IMF slopes and galaxy properties using Spearman rank coefficients, with Monte Carlo simulations to account for measurement uncertainties. For each pair of variables, we generate 5000 realisations by adding Gaussian noise based on the uncertainties of the parameters, and compute the Spearman coefficient ($\rho$) for each realisation. To simplify the notation, we report the median ($\tilde{\rho}$) of the realisations with its standard deviation ($\sigma_\rho$) and the fraction of realisations that preserve the sign of $\rho$ ($r$) in the form $\tilde{\rho} \pm \sigma_\rho \, (r\%)$. 


To avoid confusion arising from the broad use of terms and symbols in the literature (as discussed in \citealt{hopkins_2018}), we establish here the IMF-related conventions adopted throughout this work. We distinguish between the terms ``bottom'' and ``top'' IMF, referring to the mass ranges below and above $1\, \rm M_\odot$, respectively. While we nominally define the low- and high-mass IMF regimes as $0.08~M_\odot < M < 1~M_\odot$ and $1~M_\odot < M < 120~M_\odot$, respectively, the fitted slopes do not uniformly trace these full intervals. The low-mass slope is primarily constrained by TiO absorption features, which are most sensitive to M-dwarf stars in the mass range $\sim 0.08$–$0.6~M_\odot$ \citep{Henry_2024}. Similarly, the high-mass slope is mainly constrained by H$\alpha$ emission, which predominantly traces ionising stars with masses $\gtrsim 10~\rm M_\odot$. The adopted division at $1~M_\odot$ therefore serves as a practical convention, and the inferred slopes should be interpreted as effective proxies for the broader low- and high-mass IMF behaviour under the assumed power-law parametrisation.

All IMF slopes in this work are defined including the negative sign; for example, the Salpeter slope \citet{salpeter_1955} is $\alpha = -2.35$. For $\alpha_{\rm{low}}$, a more negative (``steeper'') slope indicates a larger relative number of low-mass stars compared to the Salpeter slope, whereas a more positive (``flatter'' or ``shallower'') slope corresponds to fewer stars. For the $\alpha_{\rm{high}}$ the interpretation is reversed: a slope more negative than a Salpeter slope indicates fewer high-mass stars, while a slope less negative (``shallower'') indicates a larger relative number of high-mass stars. The terms ``heavy'' and ``light'' thus refer to the relative abundance of stars in each mass regime compared to a Salpeter IMF, such that a ``bottom-heavy'' IMF has a steeper low-mass slope and a larger proportion of low-mass stars, whereas a ``bottom-light'' IMF has a flatter low-mass slope. Analogously, a ``top-heavy'' IMF has a shallower high-mass slope (more high-mass stars), and a ``top-light'' IMF has a steeper high-mass slope (fewer high-mass stars) relative to the Salpeter slope.

\subsection{The low- and high-mass end slopes}
\label{sec:results_low_high}

\begin{figure}[hbt!]
\centering
\includegraphics[width=0.9\linewidth, height=0.5\textheight, keepaspectratio]{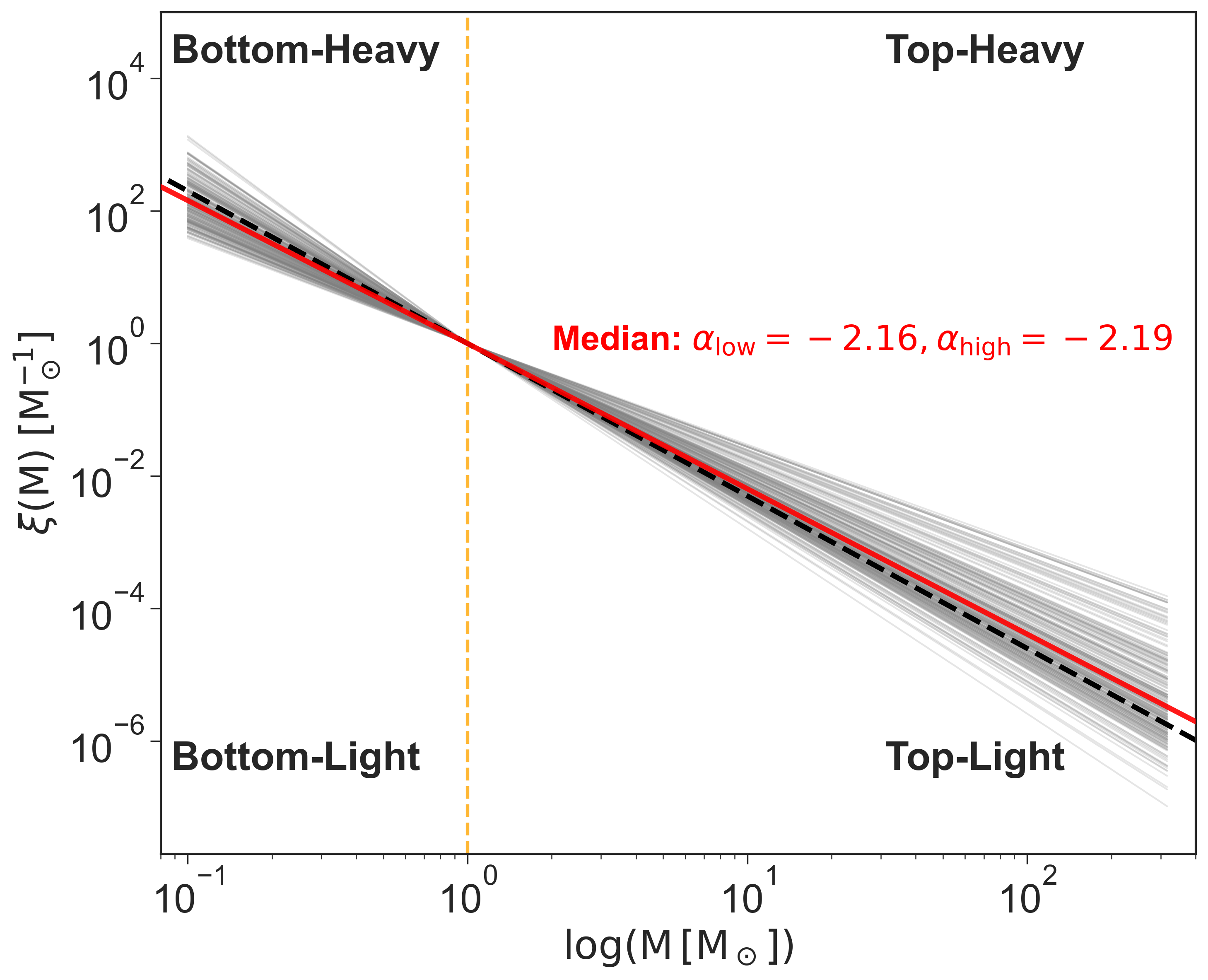}
\caption{Shapes of the IMF in logarithmic space for a representative subset of galaxies in our sample. Each grey line corresponds a different galaxy, showing the IMF slope from the low-mass end (0.08 to 1 $\rm M_\odot$) to the high-mass end (1 to 120 $\rm M_\odot$). The orange dashed vertical line indicates the transition between low- and high-mass ends. The dashed black line represents the canonical Salpeter IMF ($\alpha_{\rm{low}} = \alpha_{\rm{high}} = -2.35$). The solid red line shows the median IMF slopes across the sample ($\alpha_{\rm{low}} = -2.11$, $\alpha_{\rm{high}} = -2.19$).}
\label{fig_fullshape}
\end{figure}

Figure~\ref{fig_fullshape} illustrates the diversity of full IMF shapes in our sample, where $\alpha_{\rm{low}}$ spans approximately $-3.0 \lesssim \alpha_{\rm{low}} \lesssim -1.5$ ($\sigma \approx 0.35$), and $\alpha_{\rm{high}}$ ranges from $-2.8 \lesssim \alpha_{\rm{high}} \lesssim -1.5$ ($\sigma \approx 0.28$), suggesting that the IMF slopes vary appreciably across the sample. Across our 214 galaxies, the median values (16th–84th percentiles) are $\alpha_{\rm{low}} = -2.16$ ($-2.55$ to $-1.84$) and $\alpha_{\rm{high}} = -2.19$ ($-2.39$ to $-1.88$).

To characterise the overall IMF shape, we classify galaxies into four categories relative to the canonical Salpeter IMF ($\alpha = -2.35$): bottom-heavy/top-heavy (HH), bottom-heavy/top-light (HL), bottom-light/top-heavy (LH), and bottom-light/top-light (LL). According to this classification, 21.02\% (45) of galaxies are HH, 7.94\% (17) are HL, 47.19\% (101) are LH, and 23.83\% (51) are LL. We note that the classification into HH, HL, LH, and LL categories is purely referential and does not hold any strong physical meaning.

We find a weak but statistically robust anti-correlation, $\rho = -0.145 \pm 0.056$ (99.6\%), between the low- and high-mass IMF slopes, suggesting that bottom-heavy regions have a slight tendency to be simultaneously top-heavy.

To test whether this weak anti-correlation is driven by a third variable, we computed partial Spearman correlations while controlling for different galaxy properties. Controlling for spectral S/N yields $\rho = -0.12\pm0.06$ (98\%), indicating that S/N has a limited impact on the slope measurements. When controlling for stellar mass, the correlation becomes even weaker, $\rho = -0.07\pm 0.06$ (90\%). The same happens when controlling for $L_{\mathrm{H}\alpha}$, $\rho = -0.08 \pm 0.06$ (91\%), while controlling for metallicity effectively removes the correlation, $\rho = -0.008\pm0.06$ (50\%). These results suggest that the apparent anti-correlation between $\alpha_{\rm{low}}$ and $\alpha_{\rm high}$ is largely driven by their mutual dependence on other galaxy properties, particularly metallicity.


As further discussed in Section~\ref{sec:discussion_fullshape_full}, the weak anti-correlation between $\alpha_{\rm low}$ and $\alpha_{\rm high}$ indicates that the two ends are only loosely connected. This behaviour is consistent with a scenario in which each slope responds to different physical drivers operating on different timescales: the low-mass end reflecting time-averaged star formation, and the high-mass end tracing more recent activity.

\begin{figure*}
\centering
\includegraphics[width=0.95\linewidth]{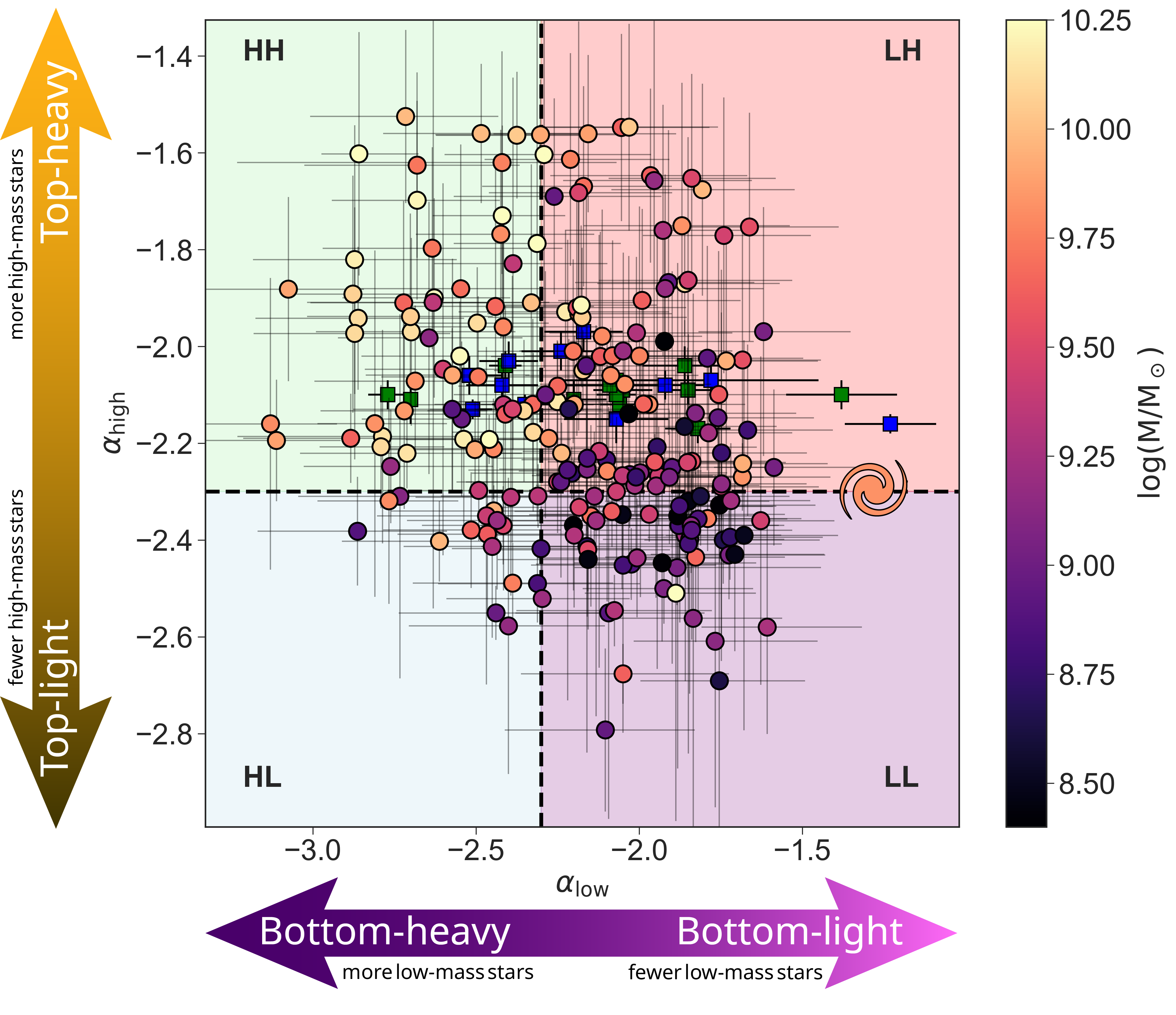}
\caption{Low-mass-end IMF slope, $\alpha_{\rm{low}}$, vs high-mass-end IMF slope, $\alpha_{\rm{high}}$, for all 214 galaxies in our sample, colour-coded by total stellar mass. The parameter space is divided into four regions: LL, LH, HL, and HH, indicating all possible combinations of low- and high-mass end slope values. The Milky Way is marked as an orange galaxy symbol, adopting the IMF from \cite{kroupa_2001}. Blue and green squares show results from \citet{denbrok_2024} for massive early-type galaxies in clusters, where green squares represent brightest cluster galaxies (BCGs) and blue squares correspond to non-BCG cluster members.}
\label{fig_quadrants}
\end{figure*}


\subsection{Links with galaxy properties}
\label{sec:results_links}


\subsubsection{Total stellar mass}
\label{sec:results_links_mass}

We first explore how our IMF slope estimates correlate with total stellar mass. Stellar mass is estimated using aperture-matched $g$- and $i$-band photometry to derive a colour–magnitude proxy following \citet{Taylor_2011}, corrected for Milky Way extinction and redshift. Notice that the stellar mass estimation assumes an IMF. Changes in the IMF shape, particularly at the low-mass end, can affect stellar masses by up to a factor of 10 \citep{Wang_2024}. However, these shifts are largely systematic, so the correlations reported here remain robust even if the absolute stellar masses are under- or over-estimated.

The low-mass slope shows a moderate negative correlation with stellar mass, $\rho = -0.25 \pm 0.06$ (99.98\%), while the high-mass slope exhibits a moderate positive correlation, $\rho = 0.27 \pm 0.06$ (100\%). This suggests that galaxies with more bottom-heavy and top-heavy IMFs tend to have higher stellar masses.

Partial correlation analysis reveals that these trends are largely insensitive to data quality: controlling for S/N yields $\rho = -0.22 \pm 0.06$ (99.9\%) for the low-mass slope and $\rho = 0.24 \pm 0.06$ (100\%) for the high-mass slope. We further test whether the observed mass trends are influenced by the well-known mass-metallicity relation (MZR; \citealt{panter_2008}): when controlling for stellar metallicity, the low-mass slope shows a weak negative trend, $\rho = -0.22 \pm 0.06$ (99.98\%), while the high-mass slope retains a moderate positive correlation, $\rho = 0.25 \pm 0.06$ (100\%). Furthermore, controlling for $L_{\mathrm{H}\alpha}$, we find a weak negative trend for the low-mass slope, $\rho = -0.16 \pm 0.06$ (99.4\%), and a weak positive trend for the high-mass slope, $\rho = 0.18 \pm 0.06$ (99.8\%). Our results indicate that the dependence of both IMF slopes on stellar mass is robust against data quality and related galaxy properties, supporting the interpretation that these trends are intrinsic rather than driven by measurement biases or selection effects.

To illustrate the mass dependence, we compared the stellar mass distributions of galaxies across different IMF slope regimes. We find that systems with steep low-mass slopes and shallow high-mass slopes (HH) are significantly more massive, with a mean stellar mass of $\log \langle \rm M_\star / M_\odot \rangle = 9.78 \pm 0.5$, than those with shallow low-mass slopes and steep high-mass slopes (LL; $\log \langle \rm M_\star / M_\odot \rangle = 9.03 \pm 0.4$). This difference is highly significant, with a KS statistic of $0.699$ ($p = 8.64 \times 10^{-12}$) and a $t$-test statistic of $7.682$ ($p = 1.56 \times 10^{-11}$). Galaxies with intermediate slope combinations (HL, LH) show comparable stellar masses ($\log \langle \rm M_\star / M_\odot \rangle \sim 9.3$–$9.5$), reinforcing the picture that extreme IMF slopes are preferentially associated with low-mass galaxies (LL) or high-mass galaxies (HH). These results indicate that IMF variation is not merely a uniform steepening in massive galaxies, but rather a mass-dependent modulation of the IMF shape: both ends of the IMF steepen in the most massive systems, suggesting a curvature that depends on stellar mass.

\begin{figure}
\centering
\includegraphics[width=0.9\linewidth, height=0.7\textheight, keepaspectratio]{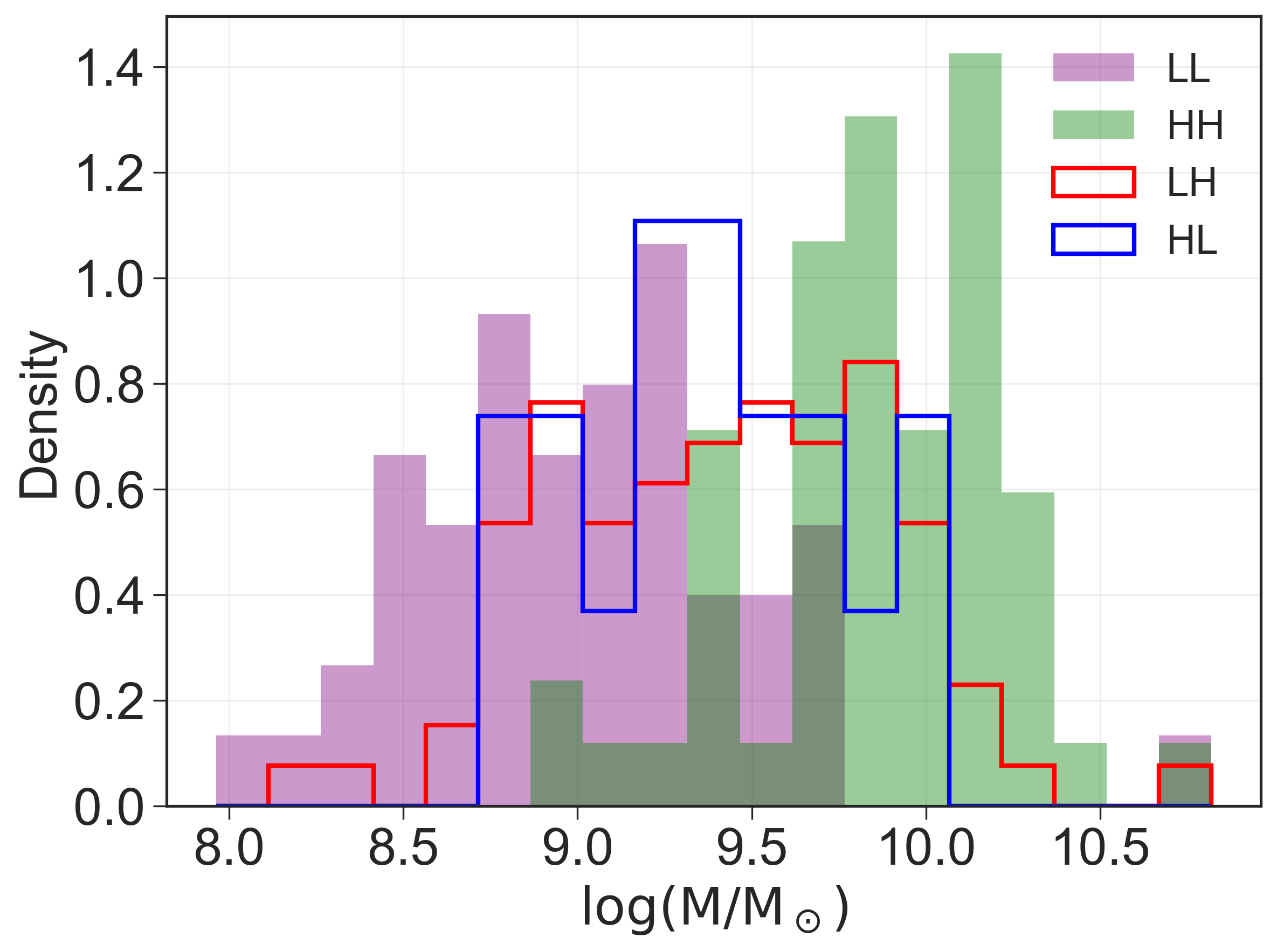}
\caption{Normalised stellar mass distributions for galaxies in each IMF quadrant. All four IMF types—LL (purple), LH (red), HL (blue), and HH (green) are well represented in the sample. LL galaxies tend to have lower stellar masses, while HH galaxies are the most massive on average. HL and LH types exhibit intermediate mass distributions with similar shapes.}
\label{fig_histograms}
\end{figure}

\subsubsection{H$\alpha$ luminosity and luminosity density}
\label{sec:results_links_sfr}

We explore how IMF slopes relate to star formation activity. In the literature, such trends are often expressed in terms of SFR and SFR surface density ($\Sigma_{\rm SFR}$), typically obtained via the \citet{kennicutt_1998} conversion, which assumes a Salpeter IMF and solar metallicity. Departures from this IMF or variations in metallicity require non-trivial corrections to recover the true SFR \citep[e.g.,][]{lee_2009, jerabkova_2018, Halsbauer_2024}. \citet{gunawardhana_2011} showed that variations in the high-mass end of the IMF have a systematic effect on the inferred SFR. However, as our analysis involves variations at both the low- and high-mass ends of the IMF, the correction of the $L(H\alpha)$–SFR relation becomes more complex. A comparison between Salpeter-SFRs and more IMF-consistent SFRs is shown in \ref{sec:appendix_SFR}. To avoid introducing strong assumptions or additional uncertainties, we therefore use the extinction-corrected $L(\mathrm{H}\alpha)$ and $L(H\alpha)$ surface density ($\Sigma_{L(H\alpha)}$) directly as proxies for star formation, and refer to these quantities as star formation indicators. This is a reasonable assumption, since H$\alpha$ emission is dominated by massive, short-lived stars ($\sim$10 Myr), and thus reliably traces recent star formation.

Following the methodology of \citet{gunawardhana_2011} and \citet{Salvador_2025}, we bin the galaxy sample in $\log(L(H\alpha))$ and $\log(\Sigma_L(H\alpha))$, and compute representative IMF slopes for each bin to characterise trends in IMF slopes. For the low-mass end, we compute the mean slope within each bin, with uncertainty given by the standard deviation. For the high-mass end, in each bin we determine the point of maximum density of the distribution of galaxies in the \lEWha\ vs $g-r$ plane via a two-dimensional kernel density estimate. We then assign the bin's slope the IMF slope corresponding to this peak, with uncertainties derived from bootstrap resampling of slope estimates.

\begin{figure*}
\centering
\includegraphics[width=0.9\linewidth]{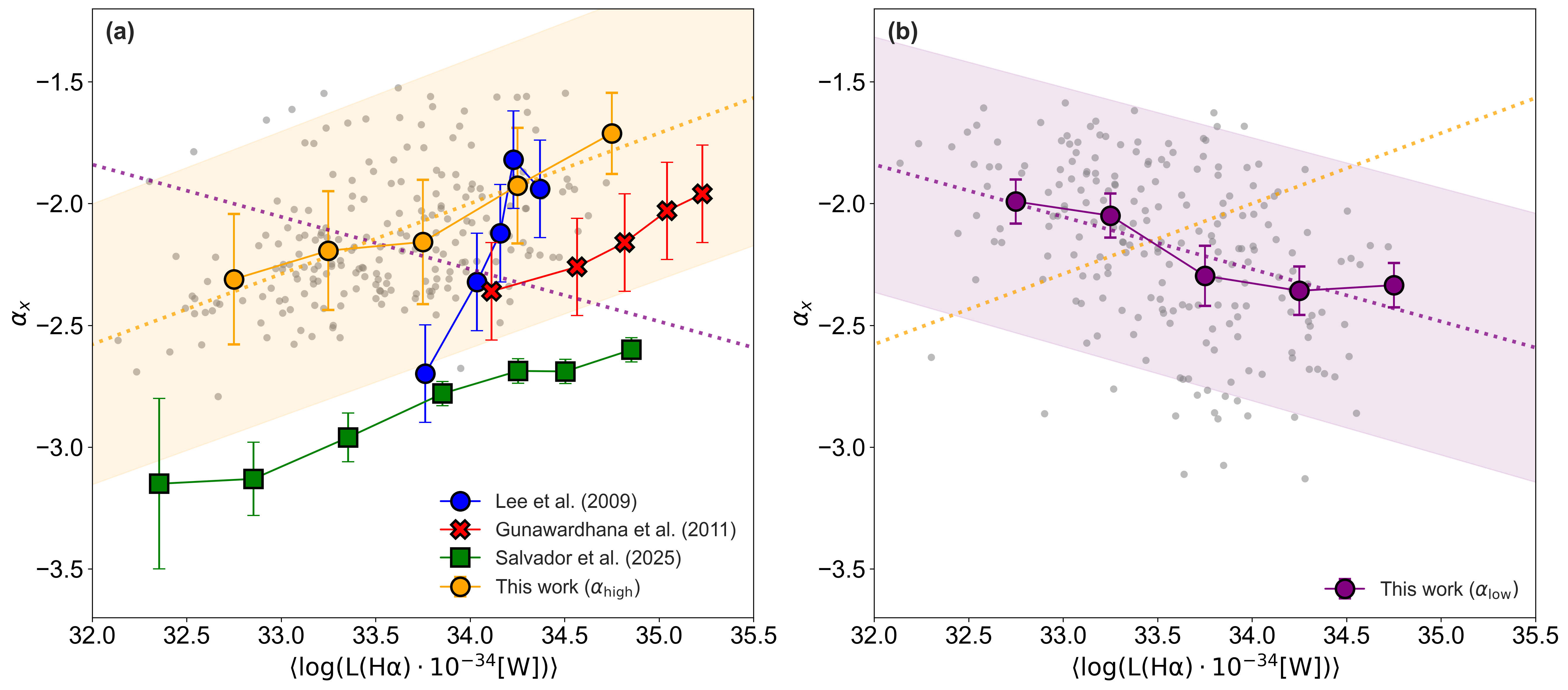}
\caption{The relation between $\alpha_X$ with$\log(L(\mathrm{H}\alpha))$, where $\alpha_X$ generically represents either $\alpha_{\rm low}$ or $\alpha_{\rm high}$. Panel a) displays the measurements of $\alpha_{\rm high}$ with orange points showing the binned values, the orange dotted line the best-fit relation, and the shaded orange region its bootstrap uncertainty. For reference, the best-fit relation for $\alpha_{\rm low}$ from panel b) is over-plotted as a purple dotted line. Panel b) shows $\alpha_{\rm low}$ with purple points and dotted line, and the $\alpha_{\rm high}$ best-fit relation from panel a) is over-plotted in orange for reference. Grey points indicate individual galaxies in both panels. Results from \citet{lee_2009} (blue), \citet{gunawardhana_2011} (red), and \citet{Salvador_2025} (green) are also shown, converted from SFR to H$\alpha$ luminosity for consistency.}
\label{fig_sfr_slope}
\end{figure*}

\begin{figure}[hbt!]
\centering
\includegraphics[width=0.9\linewidth]{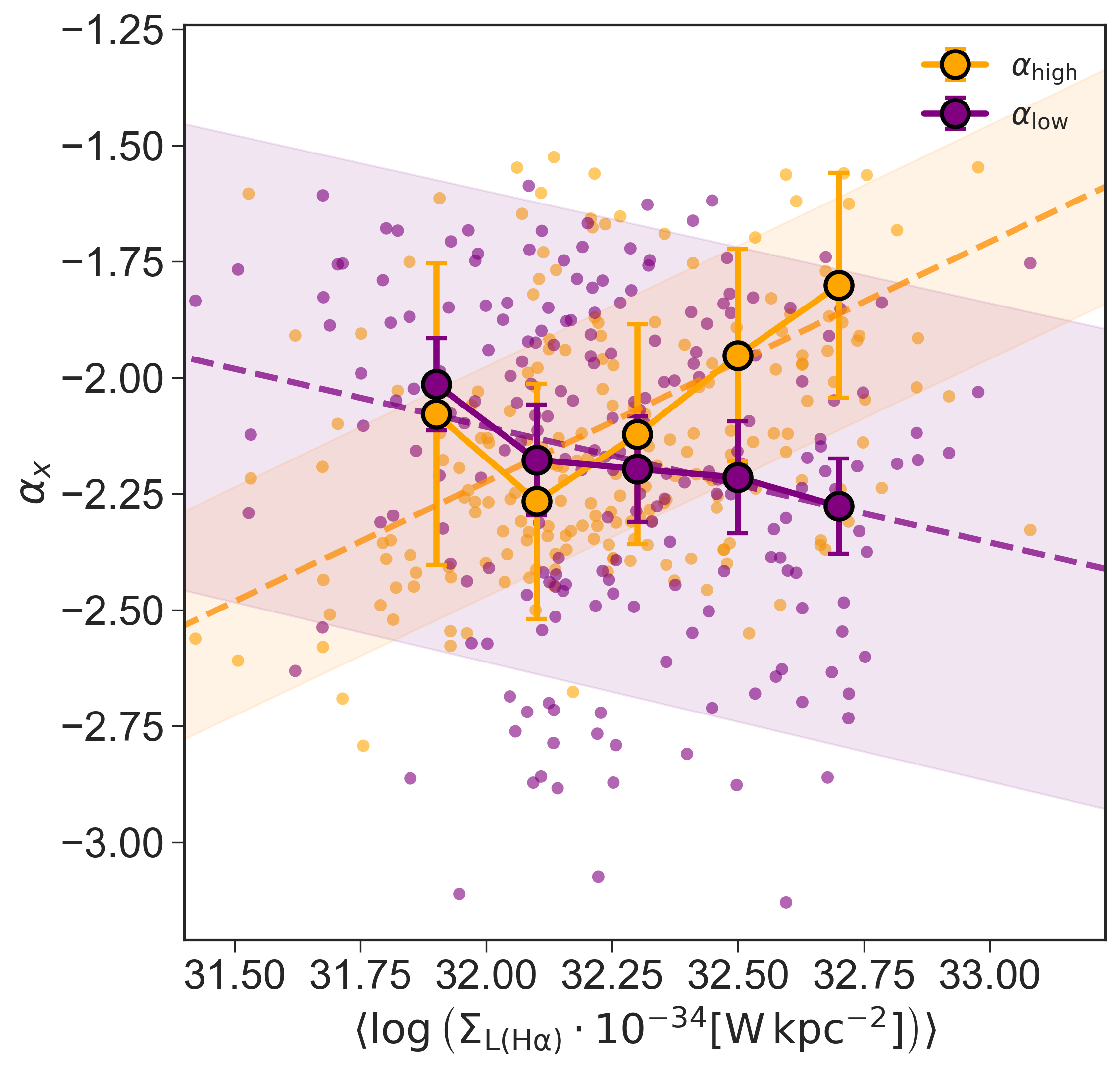}
\caption{Binned measurements of $\alpha_X$ as a function of the H$\alpha$ luminosity surface density, $\Sigma_{L(H\alpha)}$, where $\alpha_X$ generically represents either $\alpha_{\rm low}$ or $\alpha_{\rm high}$. Purple points correspond to the individual galaxy measurements for $\alpha_{\rm{low}}$, and orange points correspond to those for $\alpha_{\rm{high}}$, shown for reference. Dashed lines indicate the best-fit relations for each slope, and the shaded regions represent the corresponding bootstrap uncertainties.} 
\label{fig_sfrd_slope}
\end{figure}

We derive the following best-fit linear relations between the binned IMF slopes and $L(H\alpha)$:
\begin{align}
    \alpha_{\mathrm{low}} &= (-0.22 \pm 0.01)\,
        \langle \log(L(H\alpha) \cdot 10^{-34}\,[\mathrm{W}]) \rangle \nonumber \\
        &\quad - (5.2 \pm 0.3), \\[6pt]
    \alpha_{\mathrm{high}} &= (0.29 \pm 0.01)\,
        \langle \log(L(H\alpha) \cdot 10^{-34}\,[\mathrm{W}]) \rangle \nonumber \\
        &\quad - (12.1 \pm 0.3), \label{eq_high_SFR}
\end{align}
and $\rm{\Sigma_{L(H\alpha)}}$:
\begin{align}
    \alpha_{\mathrm{low}} &= (-0.24 \pm 0.02)\,
        \langle \log(\Sigma_{L(H\alpha)} \cdot 10^{-34}\,[\mathrm{W\,kpc^{-2}}]) \rangle \nonumber \\
        &\quad - (5.42 \pm 0.5), \\[6pt]
    \alpha_{\mathrm{high}} &= (0.51 \pm 0.04)\,
        \langle \log(\Sigma_{L(H\alpha)} \cdot 10^{-34}\,[\mathrm{W\,kpc^{-2}}]) \rangle \nonumber \\
        &\quad - (18.58 \pm 1.3). \label{eq_high_SFRD}
\end{align}
These relations are shown in Figures~\ref{fig_sfr_slope} and \ref{fig_sfrd_slope}, with uncertainties estimated via bootstrap resampling. We find negative relations between $\alpha_{\rm{low}}$ and both $L(H\alpha)$ and $\Sigma_{L(H\alpha)}$, and positive relations between $\alpha_{\rm{high}}$ and the same quantities, consistent with trends previously reported in \citet{Salvador_2025}. Our results suggest that galaxies become increasingly bottom-heavy and top-heavy with increasing $L(H\alpha)$ and $\Sigma_{L(H\alpha)}$. 

For $\alpha_{\rm low}$ we find a moderate negative correlation with $L(\mathrm{H}\alpha)$, $\rho = -0.33 \pm 0.04$ (100\%), and a weak correlation with $\Sigma_{L(\mathrm{H}\alpha)}$, $\rho = -0.11 \pm 0.05$ (97\%). When controlling for stellar mass, the correlations weaken to $\rho = -0.21 \pm 0.05$ (100\%) and $\rho = -0.10 \pm 0.05$ (97\%), respectively. Controlling for metallicity yields similar weak negative trends: $\rho = -0.21 \pm 0.05$ (100\%) and $\rho = -0.12 \pm 0.05$ (98\%).

Regarding $\alpha_{\rm high}$, we find a moderate positive correlation with $L(\mathrm{H}\alpha)$, $\rho = 0.30 \pm 0.05$ (100\%), and with $\Sigma_{L(\mathrm{H}\alpha)}$, $\rho = 0.24 \pm 0.05$ (100\%). When controlling for stellar mass, the correlations decrease to $\rho = 0.17 \pm 0.05$ (99.9\%) and $\rho = 0.24 \pm 0.05$ (100\%), respectively. When controlling for metallicity, weak but significant positive correlations remain: $\rho = 0.21 \pm 0.05$ (100\%) and $\rho = 0.26 \pm 0.05$ (100\%). Our findings suggest that stellar mass is the main driver of $\alpha_{\rm high}$, while variations in star formation activity contribute a secondary effect at a given metallicity.

Although the correlations weaken when controlling for stellar mass (and for metallicity in the case of $\alpha_{\rm low}$), the trends with recent star formation activity are still detectable. This suggests that $\alpha_{\rm low}$ and $\alpha_{\rm high}$ maintain at least a partial connection to star formation, with $\alpha_{\rm low}$ being more influenced by metallicity and $\alpha_{\rm high}$ more by stellar mass.

\subsubsection{Chemical composition}
\label{sec:results_links_chemcomp}

We examine the correlations between the IMF slopes and chemical properties, specifically [M/H] (a proxy for the total stellar metallicity) and [Mg/Fe] ($\alpha$-element enhancement tracer and commonly used as an indicator of star-formation timescales, with higher [Mg/Fe] reflecting shorter, more intense star-formation episodes), in Figure~\ref{fig_chem_slope}. For [M/H] we obtain
\begin{align}
\alpha_{\rm low}  &= (-0.40 \pm 0.13)\,\rm{[M/H]} - (2.43 \pm 0.07), \\
\alpha_{\rm high} &= (0.30 \pm 0.05)\,\rm{[M/H]} - (1.98 \pm 0.04),
\end{align}
{while for the [Mg/Fe] we find
\begin{align}
\alpha_{\rm low}  &= (0.81 \pm 0.76)\,\rm[Mg/Fe] - (2.33 \pm 0.12), \\
\alpha_{\rm high} &= (-0.54 \pm 0.48)\,\rm[Mg/Fe] - (2.07 \pm 0.11).
\end{align}
For our 214 galaxies, we find median stellar population parameters of [M/H] = –0.57 and [Mg/Fe] = 0.24, with 16th–84th percentile ranges of –0.98 to –0.20 and 0.12–0.33, respectively. We caution that individual [M/H] and [Mg/Fe] estimates may be mildly affected by residual skyline contamination (Section~\ref{sec:data_stacking}). The features most influenced by contamination are Mgb5177 and Fe5335, with median replaced pixel fractions of $\sim 8.3\%$ (16-84th: 6.8-9.7\%) and $\sim 15.6\%$ (16--84th: 14.8--22.5\%), respectively, although these remain relatively small fractions of the affected wavelength windows. Nevertheless, the preliminary consistency check in panel c) in Figure~\ref{fig_chem_slope} shows that our estimates broadly recover the expected MZR, providing confidence in the robustness of the stellar metallicities.

We find a moderate negative correlation between [M/H] and $\alpha_{\rm low}$, with $\rho = -0.41 \pm 0.04$ (100\%), and a moderate positive correlation with $\alpha_{\rm high}$, with $\rho = 0.31 \pm 0.05$ (100\%). Galaxies with higher total metallicity appear to host both steeper low-mass and flatter high-mass IMF slopes, largely independent of data quality.

Partial correlation analysis shows these trends persist when controlling for S/N: $\rho = -0.38 \pm 0.05$ (100\%) for $\alpha_{\rm low}$, and $\rho = 0.28 \pm 0.05$ (100\%) for $\alpha_{\rm high}$. A similar behaviour is seen when controlling for $L(\mathrm{H}\alpha)$: $\rho = -0.36 \pm 0.07$ (100\%) for $\alpha_{\rm low}$, and $\rho = 0.24 \pm 0.02$ (100\%) for $\alpha_{\rm high}$. Controlling for stellar mass reveals a more nuanced picture. The low-mass slope retains a significant negative correlation with metallicity, $\rho = -0.34 \pm 0.05$ (100\%), whereas the high-mass slope shows only a weak correlation, $\rho = 0.19 \pm 0.06$ (100\%). This suggests that the inferred low-mass end of the IMF correlates most strongly with stellar metallicity derived from the integrated spectra, while the high-mass end shows little dependence once stellar mass is accounted for.

Regarding [Mg/Fe], the low-mass slope exhibits a weak positive correlation, $\rho = 0.08 \pm 0.06$ (91\% of realisations positive), while the high-mass slope shows a weak negative correlation, $\rho = -0.08 \pm 0.06$ (92\% negative). Then [Mg/Fe] has only a modest influence on the IMF slopes, with higher [Mg/Fe] slightly associated with steeper $\alpha_{\rm high}$ and marginally flatter $\alpha_{\rm low}$.

Our findings are broadly consistent with those of MN24, who reported for the late-type galaxy NGC 3351 that the low-mass IMF slope becomes more bottom-heavy with increasing metallicity and more bottom-light with increasing [Mg/Fe]. Notably, MN24 found a much stronger correlation with [Mg/Fe]. This difference likely arises from the nature of the data: MN24 analysed a single galaxy with spatially resolved Voronoi bins, each with high S/N and well-constrained [Mg/Fe], covering a narrow range ($\sim$0.05–0.15). In contrast, our analysis uses global measurements for multiple galaxies spanning a broader [Mg/Fe] range (0–0.4), where the correlations with $\alpha_{\rm low}$ and $\alpha_{\rm high}$ remain weak, although slightly significant. Given the modest strength of these correlations, we therefore exclude [Mg/Fe] from our main analysis and focus instead on [M/H] as the more robust driver of IMF variations.


\begin{figure*}
\centering
\includegraphics[width=1\linewidth, keepaspectratio]{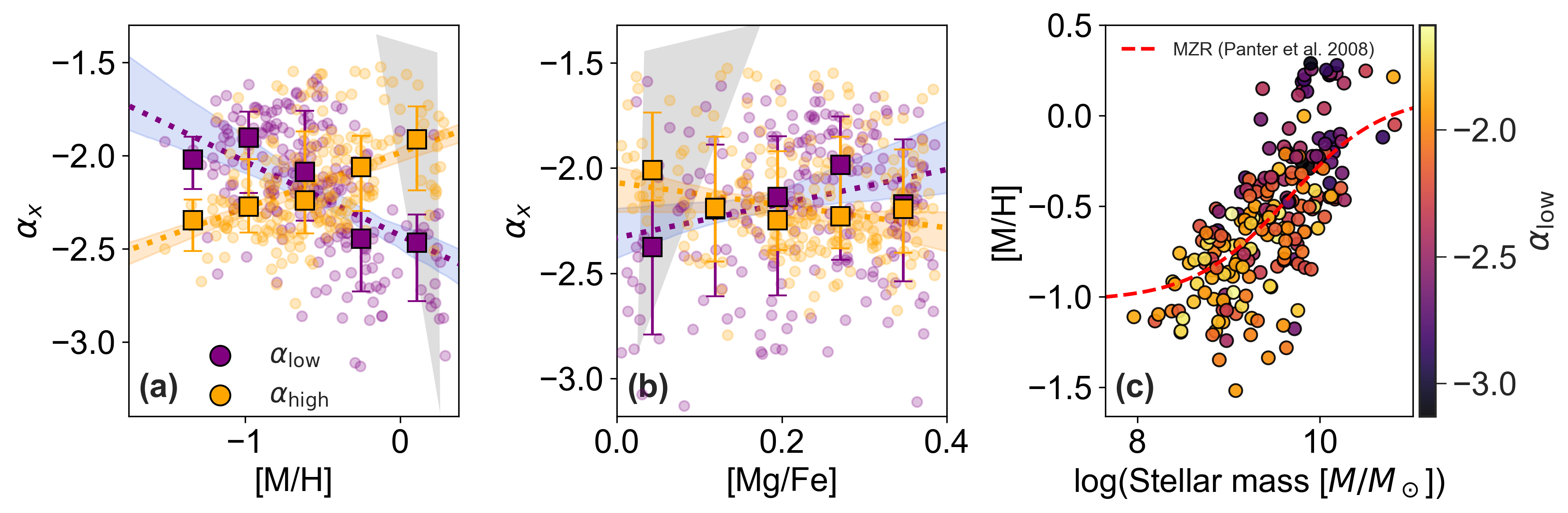}
\caption{Panels a) and b) show $\alpha_{\rm X}$ as a function of $[M/H]$ and $[Mg/Fe]$, respectively, where $\alpha_{\rm X}$ generically represents either
$\alpha_{\rm low}$ or $\alpha_{\rm high}$. The grey shaded regions in panels a)
and b) indicate the approximate location of the results from MN24, which trace
the trend of $\alpha_{\rm low}$ with chemical abundances in individual Voronoi
bins of the star-forming galaxy NGC~3351. Panel c) presents the stellar
mass-metallicity relation, colour-coded by $\alpha_{\rm low}$, with the MZR
from \citet{panter_2008} over-plotted for reference. In panels a) and b), purple
symbols indicate $\alpha_{\rm low}$, and orange symbols indicate
$\alpha_{\rm high}$. Square markers correspond to the median IMF slopes
measured in five equally spaced bins of the abundance parameter, with error
bars representing the 16th-84th percentile range within each bin. The dashed
lines show the best-fitting linear relations to the binned medians, while the
shaded regions indicate the associated uncertainties derived from 5000 bootstrap
realisations of the binned data.}
\label{fig_chem_slope}
\end{figure*}

\section{DISCUSSION}
\label{sec:discussion}

%

Our analysis aims to simultaneously constrain both the low- and high-mass slopes of the IMF in star-forming galaxies and to examine how variations in these slopes correlate with key galaxy properties. We begin by discussing the methodological limitations and potential biases, before comparing our $\alpha_{\rm low}$ and $\alpha_{\rm high}$ measurements to previous observational studies. Finally, we place our results in a broader context by contrasting them with the Milky Way IMF and the high-mass ETG sample from \citet{denbrok_2024}.

\subsection{Limitations}
\label{sec:discussion_limitations}


\subsubsection{Low-mass end slope estimation}
\label{sec:discussion_limitations_lowmass}

We first discuss the limitations of our $\alpha_{\rm low}$ estimates. Slopes are derived from MILES SPS models under the assumption of a unimodal IMF. While this assumption may not perfectly reflect the true physical IMF (MN24, Section~5.1), it provides a reasonable and practical approach given the current lack of firm constraints on the low-mass end in star forming galaxies. In the same line, other stellar population models, such as those presented in \citep{Conroy_2012b, Conroy_2018} or in the sMILES models \citep{Knowles_2021}, could potentially yield different results.

A robust characterisation of the stellar population parameters requires high-quality spectra so that subtle spectral features, such as the TiO$_1$ and TiO$_2$ molecular absorption bands (see Figure~\ref{fig_MCMC_IMF_spec}), can be reliably distinguished. While a non-negligible fraction of galaxies in our sample ($\sim 24\%$) fall below the mean S/N of $\sim119$, the majority of spectra have sufficient quality for precise measurements. In the context of the MCMC fits, we obtain a mean $\chi^2/\nu$ of 0.68. Lower-quality fits may be affected by residual skyline contamination, as well as limitations in the adopted atmosphere models and line libraries, including possible non-local thermodynamic equilibrium effects. In our spectra, sky residuals primarily impact the Fe-dominated metallicity indices Fe5270 and Fe5335, for which the median fraction of replaced pixels is $\sim 6.9\%$ (16--84th: 5.6--8.3\%) and $\sim 15.6\%$ (16--84th: 14.8--22.5\%), respectively. Importantly, the IMF-sensitive TiO features are largely unaffected, so our $\alpha_{\rm low}$ constraints are not directly driven by these regions, although some caution is warranted when interpreting individual metallicity and abundance measurements. We also note that one could, in principle, match the velocity and velocity dispersion of individual spaxels before stacking, or adopt a more detailed treatment of the variable spectral resolution in the Hector data, which could further improve the analysis.

The response functions used to model [Ti/Fe] variations are taken from \citet{conroy_2012} and are computed for a 13 Gyr stellar population. While applying these to younger systems may introduce systematic effects due to their age and wavelength dependence, we expect such effects to be sub-dominant in the IMF estimation. This is supported by the physically plausible IMF slopes and trends we recover, as well as those in \citet{navarro_2024}, who successfully applied the same framework to a Milky Way–like system.

One reason the low-mass IMF slope is challenging to constrain in star-forming systems is that massive stars dominate the light budget and can outshine the contribution from low-mass stars. However, the TiO molecular features used to trace the low-mass end originate almost exclusively from low-mass stars, primarily M dwarfs, making contamination from massive stars largely irrelevant. That said, the MCMC fitting procedure uses pseudo-continuum regions to normalise the spectra, and in this sense, a significant contribution from high-mass stars could affect the continuum: if the flux is dominated by massive stars, the normalised absorption lines would appear weaker, potentially biasing the slopes toward more bottom-light values.  The contribution of high-mass stars to the continuum, along with instrumental resolution and kinematic effects, can weaken, shallow, or broaden the absorption features of interest, potentially biasing our estimates. However, the presence of a variety of systems that do not all exhibit bottom-light IMFs suggests that these systematic effects are not dominating our measurements. 

Another limitation is that, despite allowing for an extended SFH, the method assumes a single luminosity-weighted metallicity and abundance pattern across all ages. In reality, galaxies undergo chemical enrichment and these parameters may evolve with time. A fully time-dependent treatment would be more physical, but it would introduce additional degrees of freedom that may not be sufficiently constrained by the limited number of independent IMF- and abundance-sensitive features available in the optical. In practice, however, age remains the dominant driver of spectral variations: even a small fraction of young stars can strongly affect the integrated spectrum, whereas comparable changes in metallicity or abundance ratios typically have a more subtle impact, particularly at young ages where hot stars dominate the light \citep{Vazdekis_2015}. For these reasons, adopting a single mean chemical composition provides a necessary and widely used approximation to keep the problem well-conditioned.

Finally, our inferred $\alpha_{\rm low}$ span a range very similar to that reported by \citet{denbrok_2024}, suggesting that our estimates are reasonable and that we reproduce trends seen in ETGs. Our results also reveal physically meaningful correlations: galaxies with more bottom-heavy IMFs tend to be more massive, we observe significant correlations between $\alpha_{\rm low}$ and star formation indicators, and the recovered mass–metallicity relation is consistent with expectations that more massive, metal-rich galaxies host steeper low-mass IMF slopes. These findings indicate that, although the method can be further refined, the observed global trends reflect genuine variations in the low-mass IMF across our sample. While systematic effects may influence individual measurements, the focus of this work is to demonstrate the methodology and highlight trends. The agreement with previous studies and theoretical expectations provides confidence that our results are sufficiently robust.


\subsubsection{High-mass end slope estimation}
\label{sec:discussion_limitations_highmass}

The limitations of the $\alpha_{\rm{high}}$ estimation method were discussed in detail in \citet{Salvador_2025}; we briefly summarise them here for completeness. This method relies on several assumptions inherent to the generation of stellar populations with \texttt{PÉGASE}. While \texttt{PÉGASE} allows explicit variation of the IMF, certain aspects of stellar physics, such as binaries and rotation \citep[e.g.,][]{leitherer_1999, eldridge_2012, Leitherer_2014, stanway_2016, eldridge_2017}, are not included. These processes increase the ionising photon budget by extending the lifetimes and luminosities of massive stars, producing stronger \EWha and bluer colours at fixed IMF slope. Their inclusion would likely moderate the inferred strength of IMF variations, though not alter the qualitative trends. Indeed, previous studies \citep[e.g.,][]{gunawardhana_2011, nanayakkara_2017, nanayakkara_2020}, using different synthesis tools such as \texttt{STARBURST99} \citep{leitherer_1999,Hawcroft_2025}, suggest that resulting IMF slope estimates are not strongly biased by these omissions.

Several other parameters were fixed to reduce the complexity of the models, including assumptions about photon leakage, galactic winds, IMF mass cutoffs, SFH, and metallicity. We briefly outline some of these assumptions here. We adopt a fixed upper mass cutoff of $\rm m_{\rm max} = 120~\rm M_\odot$, which does not significantly affect results when varied (see \ref{sec:appendix_uppermass}). We assume an exponentially declining SFH and a constant metallicity. While this SFH choice is appropriate for local star-forming galaxies, bursty star-formation histories can produce higher \lEWha and bias IMF estimates toward more top-heavy slopes \citep{hoversten_2008, hoversten_2010, nanayakkara_2017, nanayakkara_2020}. While galaxies can move on and off the star-forming main sequence over time, most of our sample currently lies on the sequence. As such, any past starburst episodes would mainly influence broad-band colours (e.g., g–r), whereas the H$\alpha$ emission primarily traces very recent star formation. Therefore, the inferred high-mass slope ($\alpha_{\rm high}$) reflects an effective slope under the assumed smooth SFH, with potential minor contributions from past bursts affecting the colour constraints. Finally, \citet{gunawardhana_2011} demonstrated that metallicity variations have a minor effect on the estimated slopes.

\subsection{The full IMF shape and galaxy properties}
\label{sec:discussion_fullshape}

Given the central role of the IMF in galaxy evolution, and particularly in galaxy evolution models, identifying robust relationships between observable galaxy properties and the IMF shape would be highly valuable. Such correlations could allow generic IMF assumptions to be replaced with more realistic, empirically motivated values derived from readily accessible observables.

In the following subsections, we examine the low- and high-mass slopes independently, comparing our results with previous observational studies as well as with expectations from stellar population and star formation physics. We then consider the two slopes jointly to explore systematic variations in the full IMF shape. Finally, we place our findings in a broader theoretical context by comparing them with predictions from the Integrated Galaxy-wide IMF (IGIMF; \citealt{kroupa_2003,kroupa_2013,yan_2017,jerabkova_2018}), which explicitly links IMF variations to global galaxy properties.

As mentioned, while we define the low- and high-mass IMF slopes across nominal mass ranges ($0.08-1~\rm{M_\odot}$ for $\alpha_{\rm{low}}$, and $1 -120~\rm{M_\odot}$ for $\alpha_{\rm{high}}$), the observational constraints mainly probe specific regimes: M-dwarfs for the low-mass slope ($0.08-0.6~\rm{M_\odot}$), and ionising stars for the high-mass slope ($\gtrsim 10~\rm{M_\odot}$). The inferred slopes should thus be interpreted as effective descriptions under the adopted power-law prescription, rather than uniform measurements across the full intervals.

\subsubsection{The low-mass end}
\label{sec:discussion_fullshape_lowmass}

Direct measurements of $\alpha_{\rm low}$ in star-forming galaxies remain extremely scarce. To date, the only study addressing this is MN24, who analysed the nearby late-type galaxy NGC~3351 and found low-mass IMF slopes ranging from $-3 \lesssim \alpha_{\rm low} \lesssim -1.3$, encompassing values broadly consistent with, though in some cases steeper than, the Milky Way IMF. In our sample of 214 star-forming galaxies, we recover a similar range ($-3 \lesssim \alpha_{\rm low} \lesssim -1.5$), despite using a different dataset and probing galaxy-integrated scales rather than local regions.

Several studies (e.g. \citealt{ferreras_2012,navarro_2015b,vandokkum_2017}, MN24) have reported that $\alpha_{\rm low}$ becomes more bottom-heavy with increasing metallicity, as traced by [M/H], Z, or [Fe/H]. Our results are consistent with this trend. The correlation between $\alpha_{\rm low}$ and [M/H] remains statistically significant even after controlling for stellar mass, indicating that metallicity plays a primary role in shaping the low-mass IMF. 

Complementing these findings, \citet{navarro_2021} showed that, in ETGs, $\alpha_{\rm low}$ correlates with the inferred $\Sigma_{\mathrm{SFR}}$ even at fixed metallicity, particularly in low-metallicity regions ([M/H] $\lesssim -0.1$). In our sample, $\alpha_{\rm low}$ also correlates with star formation indicators, but these relations weaken when controlling for stellar mass or metallicity, indicating that they are partly driven by the mutual dependence of these quantities on other galaxy properties. Notably, the $\Sigma_{\mathrm{SFR}}$ in \citet{navarro_2021} was inferred from [Mg/Fe] ratios, which primarily trace formation timescales rather than the instantaneous star formation directly probed in our analysis. Moreover, such inferences rely on assumptions about the SFH and chemical enrichment history, which themselves are partially degenerate with the IMF slope.

It is important to note that the observed correlations may not be strictly causal. A well-established physical connection exists between [M/H] and star formation, as higher metallicity enhances gas cooling efficiency, promoting the fragmentation of molecular clouds and, consequently, the formation of low-mass stars \citep[e.g.,][]{bate_2023,gjergo_2025}. Still, the origin of the $\alpha_{\rm low}$-[M/H] correlation remains uncertain \citep{kroupa_2024}, as there is mixed evidence in the literature regarding whether metallicity directly drives IMF variations \citep[e.g.,][]{baumgardt_2023, dickson_2023}. This uncertainty mainly arises because other environmental factors such as gas density, pressure, and stellar feedback can produce similar IMF trends, and observational biases or population mixing can obscure genuine metallicity effects. Thus, while our findings strengthen the empirical connection between metallicity and $\alpha_{\rm low}$, disentangling its physical origin will require further targeted studies.

As previously discussed, $\alpha_{\rm low}$ reflects a time-averaged low-mass end of the IMF. One possible interpretation is that older stellar populations in our sample, being more metal-rich than younger ones, could indicate that an earlier generation of massive stars enriched the ISM through nucleosynthesis and supernova feedback. This chemical enrichment may have enhanced the cooling efficiency of the ISM, favouring the formation of metal-rich, low-mass stars and thus contributing to more bottom-heavy IMFs. In this scenario, galaxies that experienced stronger or more prolonged star formation in the past would naturally appear more metal-rich and more massive today, consistent with the observed correlations between $\alpha_{\rm low}$, stellar mass, and metallicity. 

\subsubsection{The high-mass end}
\label{sec:discussion_fullshape_highmass}

Our measurements of $\alpha_{\rm{high}}$ place the galaxies in our sample within similar regions of the \lEWha\ vs $g-r$ parameter space as other star-forming galaxy samples. In particular, we find good agreement with the SAMI spaxel distribution reported by \citet{Salvador_2025} (see their Figure~4), the GAMA star-forming sample from \citet{gunawardhana_2011} (panel a) of their Figure~5), the SDSS results from \citet{hoversten_2010} (Figure~3), and the high-redshift ($1.97 < z < 2.46$) ZFIRE sample analysed by \citet{nanayakkara_2017}. All of these results, except \citet{Salvador_2025}, are $k$-corrected to $z = 0.1$, whereas our measurements are at $z \sim 0$. In all cases, the bulk of the data lie near the Salpeter slope track ($\alpha = -2.35$), with $g-r \sim 0.5$ and \lEWha $\sim 1.5-1.8$.

Quantitatively, \citet{Salvador_2025} found the following relation between $\alpha_{\rm{high}}$ and $\Sigma_{\mathrm{SFR}}$:
\begin{equation}
\alpha_{\rm high} = (0.26\pm0.08) \cdot \log \langle \Sigma_{\mathrm{SFR}} \rangle - (2.0 \pm 0.4),
\end{equation}
while \citet{gunawardhana_2011} reported:
\begin{align}
\alpha & \approx 0.36\log\langle\mathrm{SFR}\rangle - 2.6, \\
\alpha & \approx 0.3\log\langle\Sigma_{\mathrm{SFR}}\rangle - 1.7.
\end{align}
Comparing our results with previous studies is not entirely straightforward, as our relations are expressed in terms of $L(\mathrm{H}\alpha)$ and $\Sigma_{L(\mathrm{H}\alpha)}$, rather than SFR and $\Sigma_{\mathrm{SFR}}$. While the conversion between $L(\mathrm{H}\alpha)$ and SFR is direct (depending only on the adopted IMF), the transformation from $\Sigma_{L(\mathrm{H}\alpha)}$ to $\Sigma_{\mathrm{SFR}}$ additionally requires defining the spatial extent of the emitting region, which is often non-trivial and can vary across datasets. Nevertheless, because $L(\mathrm{H}\alpha)$ is a well-established tracer of recent star formation, the qualitative trends remain directly comparable.


We find that galaxies with higher $L(\mathrm{H}\alpha)$ and $\Sigma_{L(\mathrm{H}\alpha)}$ tend to exhibit flatter high-mass IMF slopes, in broad agreement with previous studies reporting that elevated star formation activity is associated with more top-heavy IMFs \citep[e.g.,][]{lee_2009,Weidner_2013,gunawardhana_2011,Salvador_2025}. Partial correlation analysis shows that these trends weaken slightly when controlling for stellar mass, but remain statistically significant, and are largely unaffected by metallicity. This indicates that the dependence of $\alpha_{\rm high}$ on H$\alpha$ luminosity and surface density is not solely driven by global galaxy parameters, but likely reflects a genuine connection with recent star formation activity. Stellar mass may still modulate $\alpha_{\rm high}$, but the intrinsic link with star-formation tracers appears robust.

Note that the relationship between stellar metallicity and recent star formation is not straightforward: our metallicity estimates trace the integrated chemical enrichment history, whereas H$\alpha$ reflects current star formation. This temporal mismatch, combined with the inter-dependencies among stellar mass, metallicity, and star-formation activity, means that partial correlations should be interpreted cautiously, as they may not fully disentangle the underlying physical connections.

Building on the latter, the positive correlation between $\alpha_{\rm high}$ and [M/H] should likewise be interpreted cautiously. As mentioned before, physically, higher metallicity promotes more efficient gas cooling, which generally favours the formation of low-mass stars. Explaining how a metal-rich ISM could give rise to more massive stars is therefore non-trivial. Massive stars preferentially form in the dense clumps within fragmenting molecular clouds, where sufficient gas mass has accumulated to support their formation. During the collapse of a molecular cloud, stars form within individual clumps. The final mass of a star is largely set by the amount of gas that can accumulate in its parent clump before collapse. This characteristic mass is approximately given by the Jeans mass ($M_J$, \citealt{Jeans}), which defines the threshold at which a clump becomes gravitationally unstable. Clumps with a higher $M_J$ can gather more mass before collapsing, producing more massive stars. The Jeans mass depends on several local conditions, such as turbulence, magnetic fields, stellar feedback, and radiation pressure—which can counteract the cooling effects of high metallicity and increase the effective temperature or pressure within the cloud \citep{larson_1992,larson_1998,larson_2005,bate_2005,bonnell_2006,grasha_2017,gouliermis_2018}. Thus, even in a metal-rich ISM, these processes may enable the formation of more massive stars by locally increasing $M_J$. While such mechanisms offer plausible explanations, the physical connection between stellar metallicity and $\alpha_{\rm high}$ remains uncertain, and our results should be regarded as indicative rather than conclusive.

\subsubsection{The full IMF shape}
\label{sec:discussion_fullshape_full}


The full shape of the IMF has been extensively studied in the Milky Way through star counts of field stars, OB associations, and globular and open clusters across both the disk and the bulge (see \citealt{elmegreen_2009}, \citealt{bastian_2010}, \citealt{hopkins_2018}, \citealt{Hennebelle_2024}, and references therein). Despite variations in physical properties among clusters, associations, and field populations, the Milky Way's IMF is generally well described by a \citet{kroupa_2001} parametrisation: $\alpha_1 \approx -1.3$ for $0.08 < \rm M/M_\odot < 0.5$, and $\alpha_2 = \alpha_3 = -2.35$ for $\rm M/M_\odot > 0.5$. 

Outside the Milky Way, however, systematic IMF variations have been reported across galaxies with different physical properties (e.g. \citealt{vanDokkum_2010, gunawardhana_2011, conroy_2012}). This apparent discrepancy may partly reflect fundamental differences in how the IMF is measured. In the Milky Way, the IMF can be directly constrained through resolved stellar counts, spanning low- to high-mass stars and diverse star-forming environments. In external galaxies, by contrast, the IMF must be inferred indirectly from integrated light, relying on population synthesis models and additional assumptions about the star formation history, chemical composition, and dust attenuation, among others. These integrated methods probe different mass ranges and star formation timescales, making them inherently non-equivalent to direct stellar counts. In Figure~\ref{fig_quadrants}, the Milky Way appears to be offset from the bulk of the distribution. This offset does not necessarily imply an inconsistency in our measurements, but likely reflects the intrinsic differences between resolved and integrated IMF estimation techniques. Galaxies with similar underlying physical conditions may therefore appear displaced because the IMF diagnostics probe different stellar populations and evolutionary timescales.

It is worth noting that adopting the Milky Way IMF as the canonical reference would shift our classification: none of the galaxies in our sample would be identified as bottom-light under that definition. The ``heavy'' and ``light'' labels are, nonetheless, purely comparative, reflecting deviations relative to an arbitrary reference slope rather than any physical transition at a specific value. There is no compelling reason to regard either the Salpeter or the Milky Way IMF as a universal benchmark. While the Milky Way may represent a system with relatively typical star formation conditions at the present epoch, it is not evident that this IMF has always been representative across different galaxy types and cosmic time.

Among the few studies attempting to constrain both the low- and high-mass ends of the IMF, \citet{denbrok_2024} analysed 25 massive ($\gtrsim 10^{12}~ \rm M_\odot$) early-type galaxies in clusters, including 14 brightest cluster galaxies (BCGs) and 11 non-BCG satellites, finding that $-3.0 \lesssim \alpha_{\rm low} \lesssim -1.3$, consistent with other ETGs studies (e.g., \citealt{ferreras_2012}; \citealt{spiniello_2014}; \citealt{vandokkum_2017}). Their high-mass slopes were inferred from chemical abundance patterns, tracing the IMF at the time of massive star formation rather than the present-day high-mass end. They found a weak and statistically insignificant correlation between $\alpha_{\rm low}$ and $\alpha_{\rm high}$, but a weakly significant one when restricting the analysis to the 11 non-BCG (satellite) galaxies, suggesting that more bottom-heavy systems may have experienced more top-heavy IMFs in the past.

Our $\alpha_{\rm low}$ measurements in LTGs span a similar range ($-2.8 \lesssim \alpha_{\rm low} \lesssim -1.3$) as those reported by \citet{denbrok_2024} in ETGs, although we probe $\alpha_{\rm low}$ in different galaxy types and environments. Both studies find a very weak correlation between the low- and high-mass slopes. In our sample, this weak anti-correlation is largely mediated by stellar mass and metallicity, while in \citet{denbrok_2024} it is unclear whether third variables play a role, although it is shown that environment could be. Importantly, \citet{denbrok_2024} estimate the high-mass slope using $\alpha$-elements, tracing past star formation, whereas we use H$\alpha$, tracing very recent star formation. Meanwhile, both $\alpha_{\rm low}$ measurements reflect longer-term averages. This difference in timescales naturally decouples the low- and high-mass ends, suggesting that the weak correlation observed does not necessarily imply the absence of a physical connection between the IMF ends, but rather reflects limitations of the current tracers in probing coeval formation at both mass regimes.

The low- and high-mass regimes could still be sequentially linked over longer, cosmological timescales. For instance, an early episode of top-heavy star formation may enrich the ISM, creating conditions conducive to a subsequent bottom-heavy IMF, as suggested by \citet{kroupa_2024}.

A clue to this link might be stellar mass, as it appears to be the only parameter that is not strongly mediated by other variables for either slope. There is a systematic trend in stellar mass across our IMF sub-samples: LL galaxies have the lowest masses, HH the highest, and LH/HL occupy intermediate values. This ordering indicates a connection between the full IMF shape and galaxy stellar mass. In this sense, stellar mass may act as an overarching parameter governing IMF variation across the full mass spectrum, even if our estimates of the low- and high-mass slopes appear decoupled when considered independently.

Interestingly, LH and HL galaxies exhibit comparable stellar mass distributions despite their opposite IMF configurations. At fixed total stellar mass, LH systems would naively be expected to contain fewer low-mass stars (and therefore less stellar mass locked into long-lived populations) than HL systems. The similarity in their stellar mass distributions may therefore point to additional processes at play. This could reflect differences in star formation history, gas accretion, or mass assembly pathways, which regulate how stellar mass builds up over time. Alternatively, it may be related to the way the IMF quadrants are defined, or to compensating trends in which more massive systems simultaneously favour enhanced formation at both ends of the IMF, as suggested by the independent correlations we observe. Disentangling these possibilities will require further investigation.

Studies of the full IMF shape remain limited, and current methods for simultaneously constraining both ends still require refinement. Future work should focus on robustly measuring low- and high-mass slopes in large, homogeneous, high-S/N, and high-resolution samples, with careful consideration of the stellar mass ranges and formation timescales traced by each indicator, as well as associated uncertainties and degeneracies. Such efforts are essential to determine whether the weak correlations observed so far are universal and to clarify the physical connection between the low- and high-mass regimes.

\subsubsection{The IGIMF framework and model predictions}
\label{sec:discussion_fullshape_IGIMF}


The Integrated Galaxy-wide IMF (IGIMF) theory \citep{kroupa_2003,kroupa_2013,yan_2017,jerabkova_2018} constructs the galaxy-wide IMF by summing the contributions of individual embedded star-forming regions over a characteristic timescale of $\delta t \sim 10$ Myr, accounting for variations in metallicity and SFR. Empirical calibrations, based on Milky Way field stars, globular clusters, and ultra-compact dwarf galaxies, parameterise the low-mass slopes $\alpha_1$ ($0.08$-$0.5 \, \rm M_\odot$) and $\alpha_2$ ($0.5$-$1 \, \rm M_\odot$), and the high-mass slope $\alpha_3$ ($1$-$150 \, \rm M_\odot$) as a function of metallicity and SFR using the empirical relations from \citet{kroupa_2002} and \citet{Marks_2012}. For a complete overview of the IGIMF framework, we refer the reader to \citet{yan_2017}, \citet{jerabkova_2018}, and \citet{yan_2019}.

Notably, IGIMF studies \citep{jerabkova_2018,Halsbauer_2024} show that the galaxy-wide IMF systematically varies with global SFR ($\psi$) and metallicity. High-SFR galaxies ($\psi \gtrsim 1\, \rm M_\odot\text{yr}^{-1}$) tend to have top-heavy IMFs, while low-SFR systems ($\psi \lesssim 1\, \rm M_\odot\text{yr}^{-1}$) are top-light. Independently of $\psi$, super-solar metallicities ([Z] $> 0$) produce bottom-heavy IMFs, sub-solar metallicities ([Z] $< 0$) yield bottom-light IMFs, and canonical IMFs occur near solar metallicity and SFR $\sim 1 \, \rm M_\odot$/yr.

Although the publicly available \texttt{galIMF} code \citep{yan_2017,yan_2019} allows for direct IGIMF predictions given a galaxy’s SFR and metallicity, such a comparison is beyond the scope of this work. This is because (1) transforming our observed $L(H\alpha)$ into SFRs would require accounting for IMF variations, which we aim to avoid, and (2) the IGIMF is constructed over a characteristic timescale of $\delta t \sim 10$ Myr for both the low- and high-mass ends, whereas our measurements probe the high-mass end on a similar $\sim 10$ Myr timescale but the low-mass end as a longer-term temporal average. Therefore, they are not directly comparable. Nonetheless, our results qualitatively align with IGIMF predictions, showing that higher metallicity correlates with a more bottom-heavy IMF and higher SFR correlates with a more top-heavy IMF.

\section{CONCLUSION}
\label{sec:conclusion}

Although the IMF is a key assumption in studies of stellar populations and galaxy evolution, its full shape in external galaxies remains largely unknown. In this work, we perform, for the first time in star-forming galaxies, a joint estimation of both the low- and high-mass ends of the IMF in 214 galaxies from the new Hector spectroscopic survey. We apply two complementary methods to the stacked spectra of star-forming spaxels in each galaxy, reaching sufficient signal-to-noise to obtain robust constraints across both ends of the IMF.

To estimate the low-mass slope ($\alpha_{\rm low}$), we adopt the SPS-based method described in MN24, fitting IMF-variable MILES SPS models to five spectral absorption features, including the TiO$_1$ and TiO$2$ molecular bands. These features are sensitive to stellar surface gravity and thus provide indirect constraints on the M-dwarf mass regime. The approach incorporates an extended star formation history, derived from 10 \texttt{pPXF} realisations within an MCMC framework for stellar population parameters. This allows us to mitigate the age–IMF degeneracy that has historically complicated studies of the low-mass regime, particularly in systems dominated by young massive stars. This method effectively probes the $\sim 0.08$–$0.6~M\odot$ range and provides a time-averaged characterisation of the low-mass IMF.

The high-mass slope ($\alpha_{\rm high}$) is inferred using the Kennicutt diagnostic, by comparing the observed \lEWha\ and $g-r$ colour with a grid of \texttt{PÉGASE} SPS models, each corresponding to a different high-mass slope. This method primarily probes the recent ($\lesssim 10~\rm Myr$) star formation activity of massive ($\gtrsim 10~M_\odot$) stars.

By combining these independent methods, we reconstruct the full IMF shape for each galaxy and examine its correlations with stellar mass, H$\alpha$ luminosity ($L(H\alpha)$) and luminosity density ($\Sigma_{L(H\alpha)}$) as indicators of star formation, as well as with stellar metallicity ([M/H]). Our main findings are summarised as follows:

\begin{itemize}
    \item We find a broad diversity of IMF shapes across our sample. For clarity, we separate galaxies into four categories based on the combination of their low- and high-mass slopes (with respect to the Salpeter slope): bottom-heavy/top-heavy (HH), bottom-heavy/top-light (HL), bottom-light/top-heavy (LH), and bottom-light/top-light (LL). According to this classification, 21.02\% (45) of galaxies are HH, 7.94\% (17) are HL, 47.19\% (101) are LH, and 23.83\% (51) are LL. While this division is arbitrary, it highlights that the low- and high-mass ends can vary independently and span a wide range of combinations.

    \item We find a weak but statistically significant anti-correlation between $\alpha_{\rm low}$ and $\alpha_{\rm high}$ ($\rho = -0.145$). However, partial correlation analysis shows that this trend weakens substantially when controlling for stellar mass and $L_{\mathrm{H}\alpha}$, and disappears entirely when controlling for metallicity. As $\alpha_{\rm low}$ and $\alpha_{\rm high}$ probe different star formation timescales, although the observed anti-correlation is weak and largely explained by secondary dependencies, it does not necessarily rule out a physical connection between both ends of the IMF. Instead, it may reflect the distinct physical drivers and temporal sensitivity of the tracers used to constrain each slope.

    \item We find that $\alpha_{\rm low}$ exhibits a moderate negative correlation with stellar mass, while $\alpha_{\rm high}$ shows a moderate positive correlation. These trends remain largely unchanged when controlling galaxy properties or data quality, indicating that the mass dependence is intrinsic rather than driven by secondary correlations. More massive galaxies tend to be simultaneously bottom- and top-heavy, consistent with HH systems being the most massive and LL systems the least massive in our sample. Galaxies with mixed slope combinations occupy intermediate stellar masses. This systematic progression suggests that IMF variation is not simply a uniform steepening with increasing mass, but rather a mass-dependent modulation of the full IMF shape, where both ends become steeper in the most massive galaxies.

    \item $\alpha_{\rm low}$ correlates negatively and $\alpha_{\rm high}$ positively with both $L(H\alpha)$ and $\Sigma_{L(H\alpha)}$: high star-formation intensity is associated with both more top- and bottom-heavy IMF slopes, extending previous high-mass results to the full IMF. The low-mass trends weaken when controlling for stellar mass and metallicity, while high-mass trends only weaken with stellar mass.

    \item $\alpha_{\rm low}$ correlates negatively and $\alpha_{\rm high}$ positively with [M/H], such that higher metallicity is linked to both more bottom- and top-heavy IMF slopes. These trends are mostly independent of H$\alpha$ tracers and data quality. Controlling for stellar mass, the $\alpha_{\rm low}$–[M/H] correlation remains strong, while the $\alpha_{\rm high}$ trend weakens, suggesting that low-mass variations are driven mainly by metallicity, whereas high-mass variations depend on both metallicity and stellar mass.
    
    \item We find strong evidence supporting that IMF variations cannot be described by a single-parameter dependence. Instead, the low- and high-mass ends respond differently to multiple galaxy properties and timescales, indicating that IMF variation is intrinsically multi-dimensional.

\end{itemize}

Despite inherent uncertainties in both our observational data and the methods used to estimate the IMF slopes, we have carefully assessed and quantified the main sources of bias. While the precise values, particularly for $\alpha_{\rm low}$, remain subject to these uncertainties, the robustness of our results is supported by several factors: we broadly recover well-established scaling relations such as the mass–metallicity relation and the star-forming main sequence, and the correlations we find between IMF slopes and galaxy properties are consistent with previous observational studies and theoretical expectations. These findings demonstrate that, despite systematic limitations, it is feasible to reconstruct the full IMF in star-forming galaxies with the Hector survey.

Looking forward, further refinement of IMF measurement techniques will be crucial to fully characterise the IMF shape and its evolution over time. This requires a better understanding of IMF tracers: the stellar mass ranges and temporal scales they probe, their intrinsic degeneracies, and how to optimally combine information across different wavelengths. In this context, our work, together with previous studies, lays the groundwork for a continued, detailed, and empirical exploration of the full IMF across diverse galaxy populations.

Given the numerous studies demonstrating variability in the IMF, upcoming surveys, particularly those targeting high-redshift galaxies, must account for these variations to avoid biases in the inferred stellar masses, star formation rates, and other galaxy properties. While some efforts have already been made to explore these effects, the precise impact of specific IMF slope variations remains uncertain, especially when considering the simultaneous variation of both ends of the IMF. Galaxy evolution and stellar population synthesis models therefore need to incorporate flexible IMF parametrisations, allowing for different IMF functional forms and slopes. Doing so will be essential for constructing a self-consistent, IMF-informed picture of galaxy evolution across cosmic time. The full Hector survey, with integral-field spectroscopy of $\sim$15,000 galaxies, together with next-generation facilities such as JWST, the Extremely Large Telescope (ELT), and Euclid, will provide the data necessary to robustly test these variations across environments, galaxy types, and redshifts, ultimately improving our understanding of the cosmic evolution of the IMF.

\begin{acknowledgement}
The Hector Galaxy Survey is based on observations made at the Anglo-Australian Telescope. We acknowledge the traditional owners of the land on which the AAT stands, the Gamilaraay people, and pay our respects to elders past and present. The Hector multi-object integral field spectrograph instrument was built jointly by the University of Sydney and Macquarie University nodes of the Astralis Astronomical Instrumentation Consortium (\url{https://astralis.org.au/}) , with additional financial contributions from the Australian National University and University of Western Australia and supported by the Australian Research Council through grants LE170100242, LE190100018 and FT180100231. The Hector input catalogue is based on data taken from the WAVES Survey, Sloan Digital Sky Survey, GAMA Survey, 2dFGRS and Skymapper Southern Sky Survey. The Hector Galaxy Survey research was supported by the Australian Research Council Centre of Excellence for All Sky Astrophysics in 3 Dimensions (ASTRO 3D), through project number CE170100013, and other participating institutions. The Hector Galaxy Survey website is \url{https://Hector.survey.org.au/}. The Hector Galaxy Survey makes use of Data Central services (\url{datacentral.org.au}).

DS was supported by the Commonwealth through an Australian Government Research Training Program Scholarship (\url{https://doi.org/10.82133/C42F-K220}) to conduct this research.

AR recognises the support from the Australian Research Council Centre of Excellence in Optical Microcombs for Breakthrough Science (project number CE230100006), funded by the Australian Government. 

CF is the recipient of an Australian Research Council Future Fellowship (project number FT210100168) funded by the Australian Government. CF is a recipient of ARC Discovery Project DP210101945.

IM acknowledges support from grant PID2022-140869NB-I00 from the Spanish Ministry of Science and Innovation.

KG acknowledges support from Australian Research Council Laureate Fellowship FL180100060.

KO acknowledges support from the Korea Astronomy and Space Science Institute under the R\&D program (Project No. 2025-1-831-01), supervised by the Korea AeroSpace Administration, and the National Research Foundation of Korea (NRF) grant funded by the Korea government (MSIT) (RS-2025-00553982).

SMS acknowledges funding from the Australian Research Council (DE220100003).
Parts of this research were conducted by the Australian Research Council Centre of Excellence for All Sky Astrophysics in 3 Dimensions (ASTRO 3D), through project number CE170100013.

SO acknowledges support from the Korean National Research Foundation (NRF) (RS-2023- 00214057; RS-2025-00514475)

ST acknowledges the support from the Royal Thai Government Scholarship and the University of Sydney Postgraduate Research Supplementary Scholarship.

This work was supported by the Korea Astronomy and Space Science Institute under the R\&D program (Project No. 2025-1-831-01) supervised by the Ministry of Science and ICT (MSIT). MP acknowledges support from the National Research Foundation of Korea (NRF) grant funded by the Korea government (MSIT) (No. 2022R1A2C1004025).

This work made use of the AI tool ChatGPT (\url{https://chatgpt.com/}) for language editing and to improve readability. All scientific content, analyses, and conclusions are entirely the authors’ own.

\end{acknowledgement}

\bibliography{bibtemplate}

\appendix

\section{Robustness Checks}
\label{sec:appendix_robustness}

\subsection{Effect of [Ti/Fe] Prior on $\alpha_{\rm{low}}$}
\label{sec:appendix_TiFe}

To assess the robustness of our $\alpha_{\rm{low}}$ estimates against the choice of [Ti/Fe] prior, we re–ran the MCMC scheme described in Section~\ref{sec:method_lowmass} using six Gaussian priors centred at $-0.25$, $-0.15$, $-0.05$, $+0.05$, $+0.15$, and $+0.25$, spanning the full range of [Ti/Fe] covered by the MILES models ($-0.3$ to $+0.3$).


Figure~\ref{fig_tife_priors} shows the violin+box plot of the distribution of 
$\Delta \alpha = \alpha_{\rm{low}}\text{([Ti/Fe]=X)} - \alpha_{\rm{low}}\text{([Ti/Fe]=0)}$, 
for $X=$[$-0.25$, $-0.15$, $-0.05$, $0.05$, $0.15$, $0.25$]. 
The distributions are all sharply peaked around zero, with a global mean 
$\langle \Delta \alpha \rangle = 0.028$ and a small interquartile scatter 
(IQR $\simeq$ 0.074, corresponding to a robust dispersion of $\sigma_{\rm robust}=\text{IQR}/1.349 \simeq$ 0.055). This demonstrates that $\alpha_{\rm{low}}$ is robust to reasonable variations in the [Ti/Fe] prior and that this assumption does not introduce a significant bias.
    
\begin{figure}[h]
\centering
\includegraphics[width=0.8\linewidth]{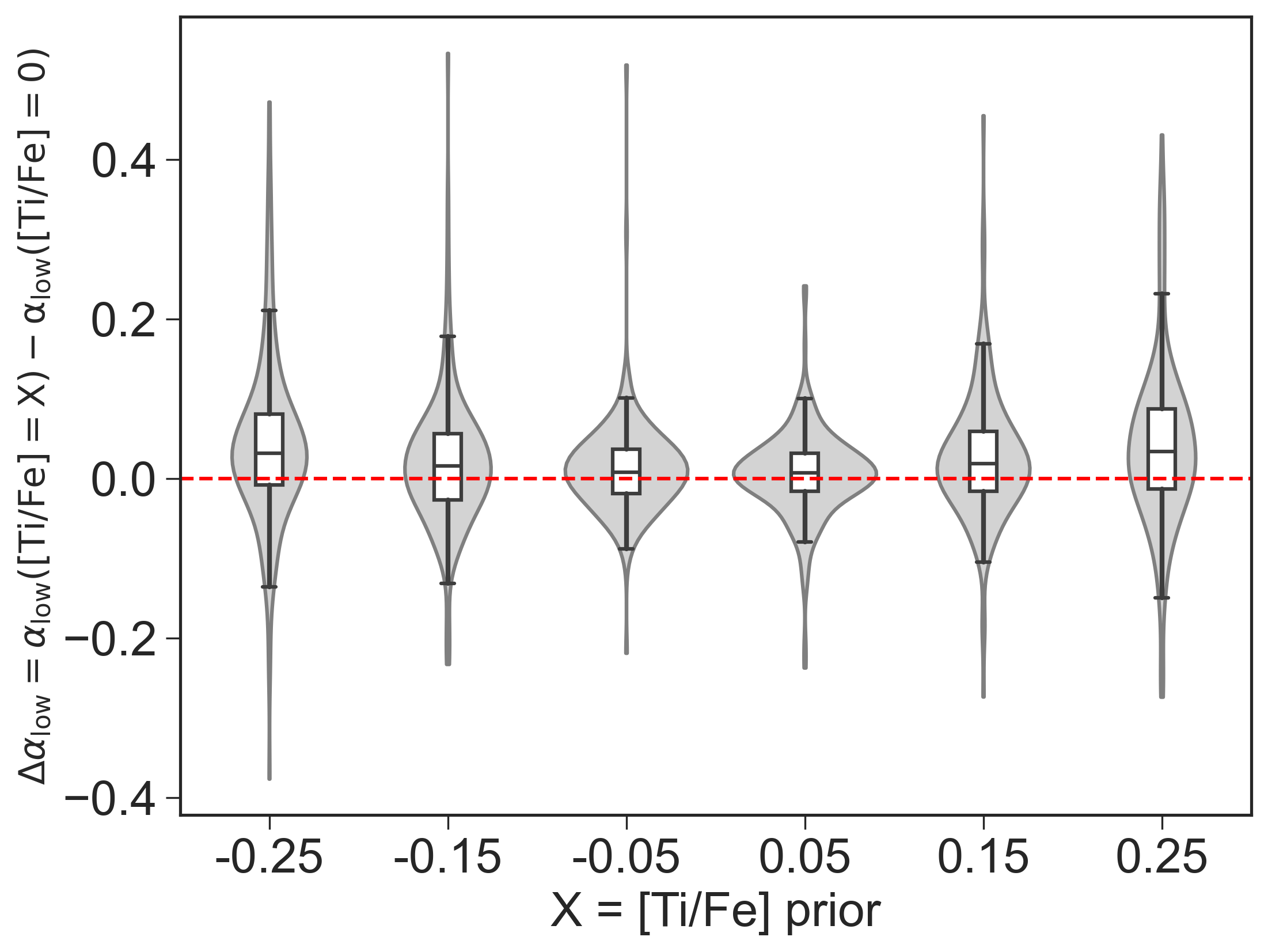}
\caption{Violin and box plot showing the distributions of $\Delta \alpha_{\rm low} = \alpha_{\rm{low}}([\rm{Ti/Fe}]=X)$ $-$ $\alpha_{\rm{low}}([\rm{Ti/Fe}]=0)$ for six different Gaussian priors on [Ti/Fe] ($X = -0.25, -0.15, -0.05, +0.05, +0.15, +0.25$). The distributions are peaked around zero, indicating that the inferred $\alpha_{\rm{low}}$ is largely insensitive to the choice of [Ti/Fe] prior and that this assumption does not introduce a significant bias.}
\label{fig_tife_priors}
\end{figure}

\subsection{Effect of NaD and TiO molecules on $\alpha_{\rm{low}}$}
\label{sec:appendix_NaD}

We investigate the impact of ISM-sensitive and IMF-sensitive spectral features on the estimation of $\alpha_{\rm{low}}$. First, we assess the effect of including the NaD line, which is known to be strongly affected by interstellar medium absorption. Second, we evaluate the IMF-constraining power of the TiO molecular bands by removing TiO$_1$ and TiO$_2$ individually from the MCMC fitting procedure to determine how strongly each feature influences the derived IMF slope.

To quantify these differences, we define $\Delta = \alpha_{\rm{low}}^{\rm{with}} - \alpha_{\rm{low}}^{\rm{without}}$ and $\alpha_{\rm w/wo} = \alpha_{\rm{low}}^{\rm{with}} / \alpha_{\rm{low}}^{\rm{without}}$.

Figure~\ref{fig_NaD} summaries the results. For NaD, we find a mean difference $\Delta_{\rm mean} = 0.246$ with a scatter $\sigma = 0.668$, and a mean ratio $\alpha_{\rm w/wo} = 0.954$ with $\sigma = 0.271$. For TiO$_1$, $\Delta_{\rm mean} = 0.210$, $\sigma = 0.618$, and $\alpha_{\rm w/wo} = 0.958$ with $\sigma = 0.251$. For TiO$_2$, $\Delta_{\rm mean} = 0.154$, $\sigma = 0.608$, and $\alpha_{\rm w/wo} = 0.977$ with $\sigma = 0.244$. These results indicate that while each feature contributes to constraining $\alpha_{\rm{low}}$, the overall estimates are relatively robust to the inclusion or exclusion of individual lines.

Although there is scatter in all cases, the differences are systematic. Including the NaD line produces the largest changes, generally driving $\alpha_{\rm{low}}$ toward steeper values, i.e., making galaxies appear more bottom-heavy, likely due to the sensitivity of NaD to ISM contamination. Removing TiO$_1$ or TiO$_2$ also decreases $\alpha_{\rm{low}}$ on average, although the effect is less pronounced. Many galaxies yield similar results with or without these features, indicating that TiO$_1$ and TiO$_2$ individually provide robust constraints on the IMF. Therefore, combining both features strengthens the $\alpha_{\rm{low}}$ constraints, enabling a more robust and reliable determination of the low-mass IMF slope.

\begin{figure}[h]
\centering
\includegraphics[width=0.65\linewidth]{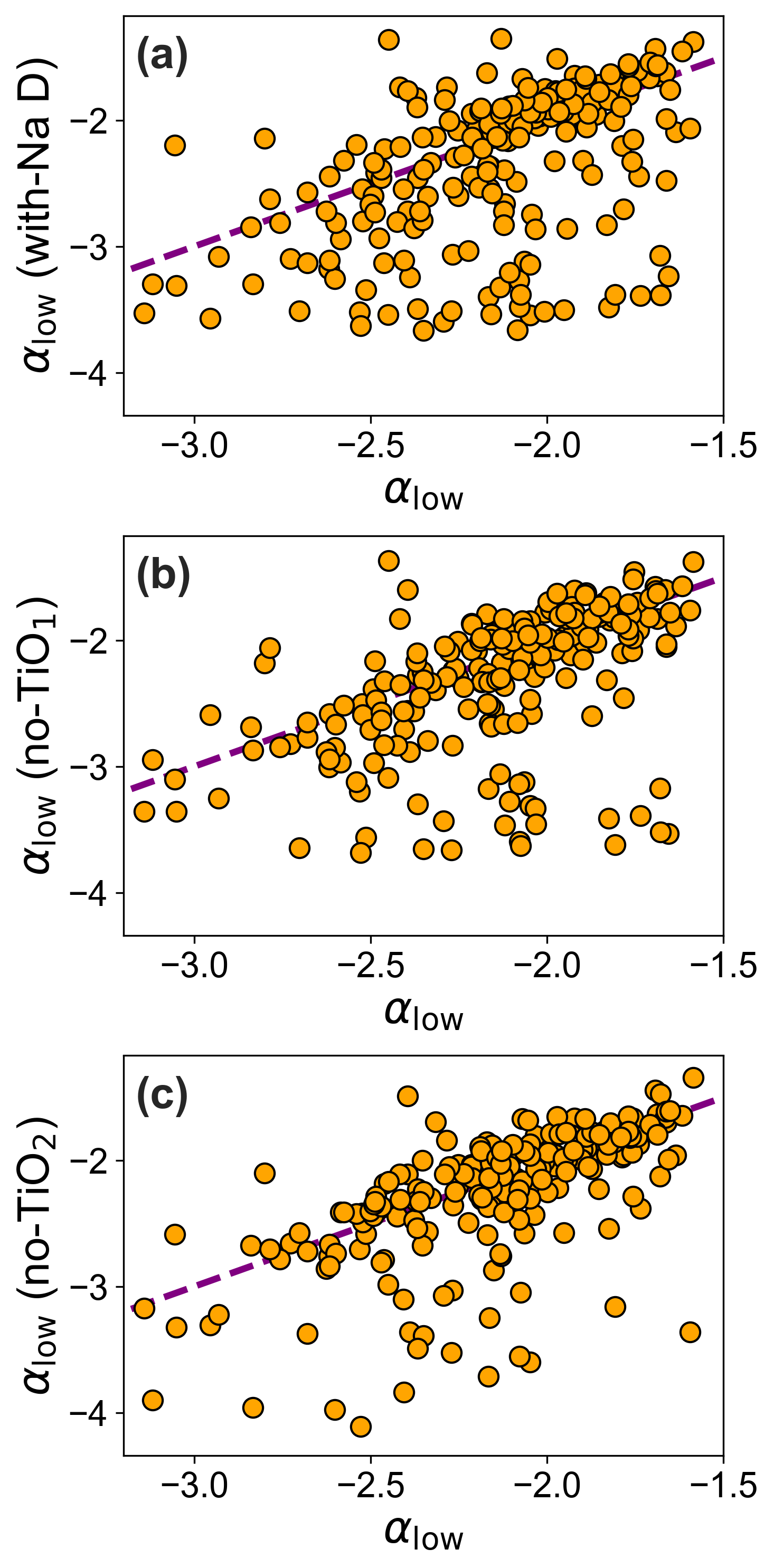}
\caption{Comparison of $\alpha_{\rm{low}}$ derived with and without specific spectral features included in the MCMC fitting. Panel a) shows results with and without NaD, panel b) with and without TiO$_1$, and panel c) with and without TiO$_2$. Orange points correspond to individual galaxies, and the purple line indicates the one-to-one relation. Overall, including NaD or omitting TiO features systematically shifts $\alpha{\rm{low}}$ toward more negative values, corresponding to a more bottom-heavy IMF.}
\label{fig_NaD}
\end{figure}

\subsection{Consistency with the IMF–velocity dispersion relation in ETGs} 
\label{sec:appendix_QS}

The MN24 method extends the techniques commonly applied to quiescent galaxies to systems with ongoing star formation (e.g., \citealt{conroy_2012}). The key improvement is the inclusion of the extended SFH in the MCMC module, which allows the model to account for more complex multi-age stellar populations. To verify that this extension yields consistent results with traditional methods, we apply the MN24 approach to stacked spectra of spaxels with no ongoing star formation from 94 quiescent galaxies in the Hector survey. 

As shown in Figure~\ref{fig_QS}, the inferred $\alpha_{\rm low}$ values are broadly consistent with the trends reported by \citet{ferreras_2012} and \citet{Labarbera_2013}, in the sense that galaxies with higher velocity dispersion tend to show steeper low-mass IMF slopes. We note, however, that the scatter is substantial, particularly at high velocity dispersion where the number of objects is small, and that at $\sigma \lesssim 150\,\mathrm{km\,s^{-1}}$ our $\alpha_{\rm low}$ values are systematically lower than those reported in the literature. Given these limitations, this comparison should be interpreted as a qualitative consistency check rather than a precise recovery of the IMF–velocity dispersion relation.

\begin{figure}[h]
\centering
\includegraphics[width=0.9\linewidth]{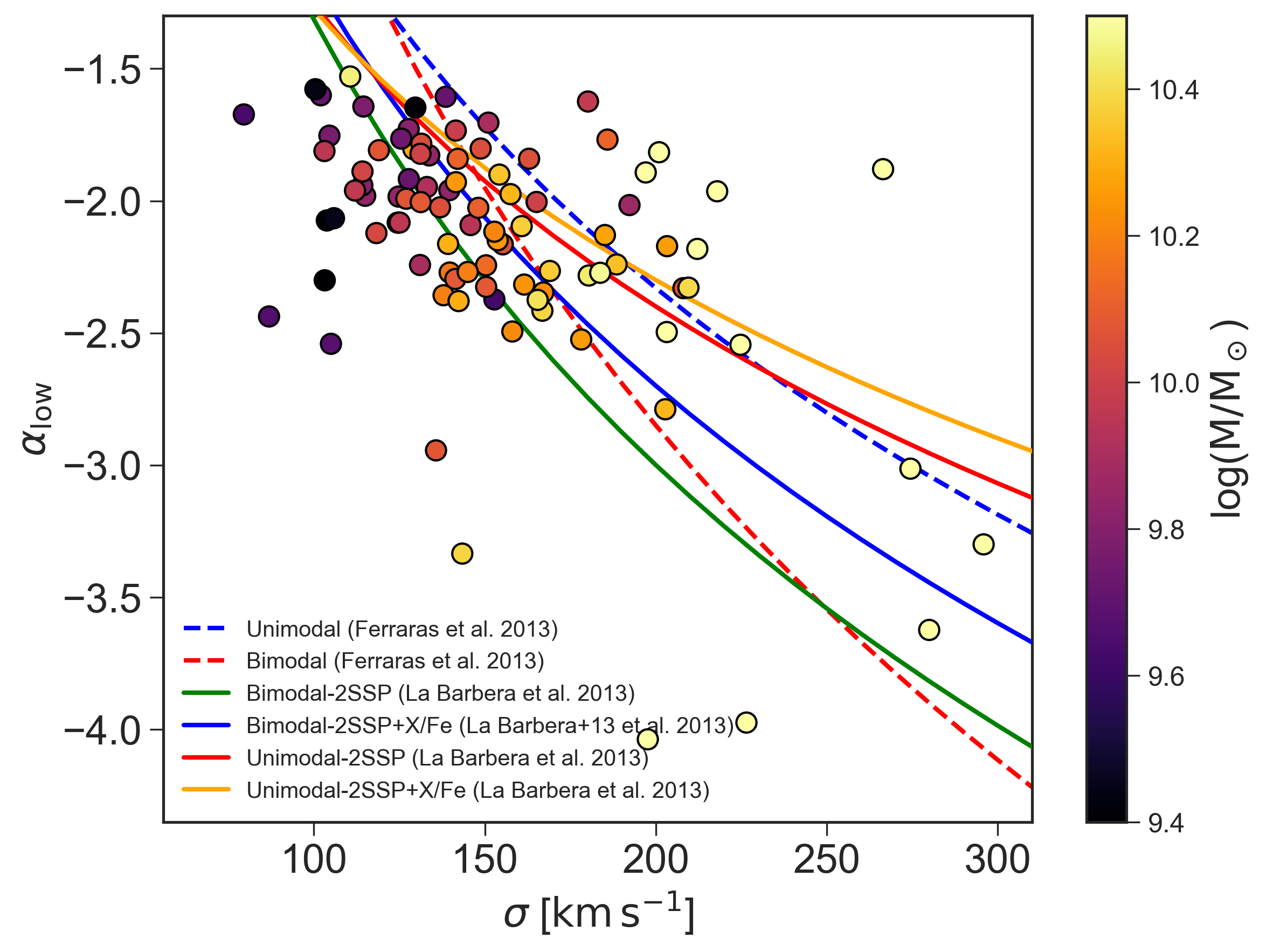}
\caption{Relation between $\alpha_{\rm low}$ and the stellar velocity dispersion ($\sigma$) for the quiescent galaxy sample. Each point represents an individual galaxy, colour-coded by stellar mass. The dashed blue and red lines show the unimodal and bimodal IMF–$\sigma$ relations from \citet{ferreras_2012}, respectively. The solid lines correspond to the relations from \citet{Labarbera_2013}: bimodal 2SSP (green), bimodal 2SSP+X/Fe (blue), unimodal 2SSP (red), and unimodal 2SSP+X/Fe (orange), where the \textit{+X/Fe} models include additional parameters for element abundance variations ([Ca/Fe], [Na/Fe], [Ti/Fe]).}
\label{fig_QS}
\end{figure}

\subsection{Effect of the Upper Stellar Mass Limit $\rm{m_{max}}$} 
\label{sec:appendix_uppermass}

We perform a check to test the effect of one parameter of our \texttt{PÉGASE} models on the resulting IMFs: the upper stellar mass limit $\rm{m_{max}}$, which sets the maximum stellar mass that can be formed in a population. Changing this parameter can affect the IMF because it controls the contribution of the most massive stars to the integrated light. In our base models we adopt $\rm{m_{max}}=120~M_\odot$. To test its impact, we compare the retrieved IMF slopes for $\rm{m_{max}}=100~M_\odot$ and $150~\rm M_\odot$.

The top and bottom panels of Figure~\ref{fig_mmax} show the difference in recovered IMF slopes relative to the fiducial $\rm{m_{max}}=120~M_\odot$ model. We find that lowering $\rm{m_{max}}$ to $100~\rm M_\odot$ produces systematically steeper $\alpha_{\rm{high}}$, with a mean offset of $\Delta\alpha_{\rm{high}}=+0.07$ and a standard deviation of $\sigma_{\Delta}=0.03$. Increasing $\rm{m_{max}}$ to $150~\rm M_\odot$ yields negligible changes, with a mean $\Delta\alpha_{\rm{high}}=0.01$ and $\sigma_{\Delta}=0.001$. These results indicate that the choice of $\rm{m_{max}}$ has only a minor impact on the inferred IMF slopes.

\begin{figure}
\centering
\includegraphics[width=1\linewidth]{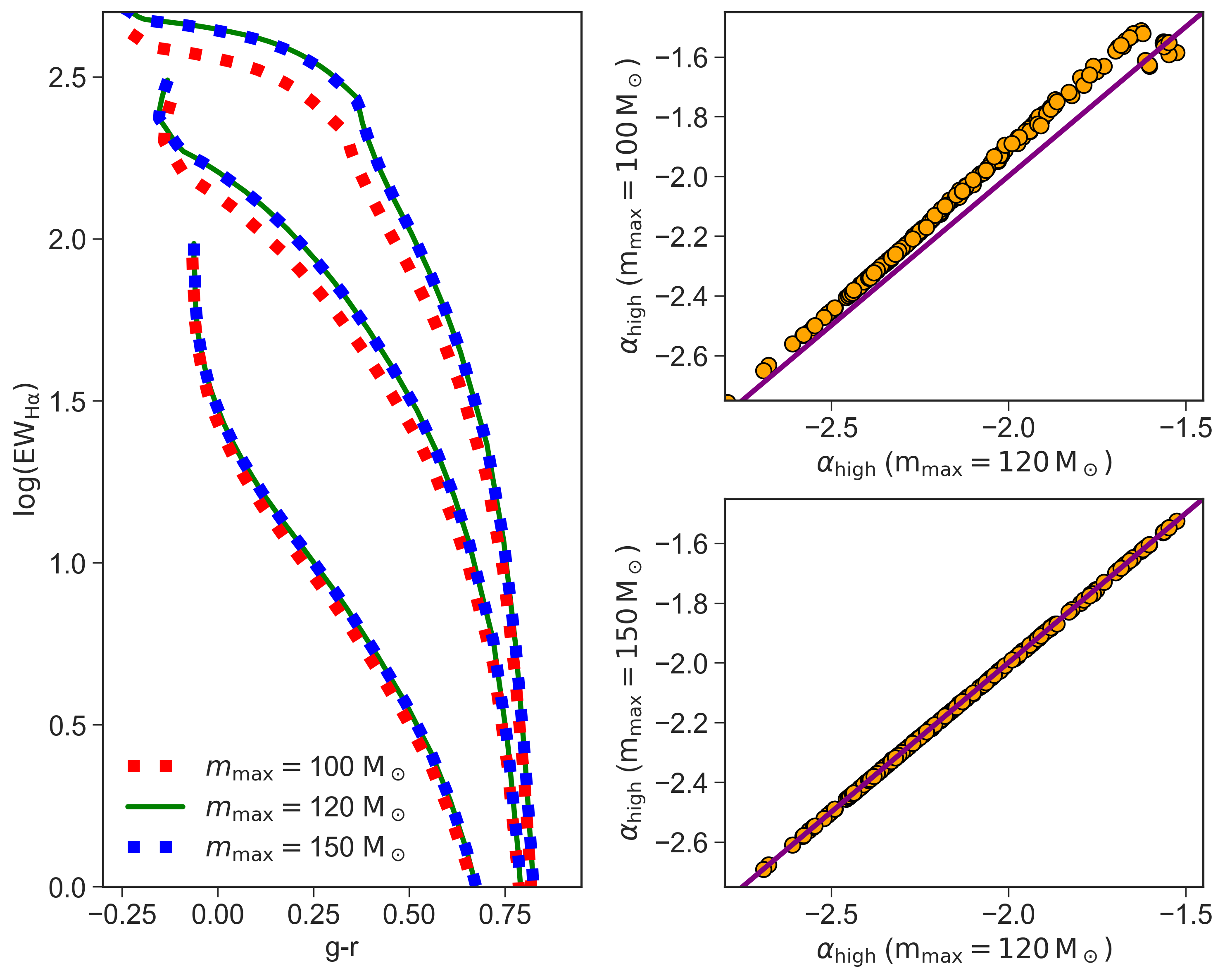}
\caption{Effect of the upper stellar mass limit ($\rm{m_{max}}$) parameter on \texttt{PÉGASE} on our results. Left: Model grids in the $\log(\rm{EW}{H\alpha})$ vs $g-r$ plane for $\rm{m{max}}=100 \, \rm M\odot$ (red dashed), $120 \, \rm M\odot$ (green solid), and $150~\rm M_\odot$ (blue dashed). Right: Differences in the recovered IMF slopes relative to the fiducial $\rm{m_{max}}=120~M_\odot$ model, shown for $\rm{m_{max}}=100~M_\odot$ (top) and $\rm{m_{max}}=150~M_\odot$ (bottom). Orange points correspond to individual galaxies.}
\label{fig_mmax}
\end{figure}

\section{IMF-consistent star formation rates}
\label{sec:appendix_SFR}

The SFRs reported in the literature are typically derived using the Kennicutt relation, which assumes a Salpeter IMF \citep{kennicutt_1989, kennicutt_1994}. Since we find variations in both ends of the IMF across our galaxy sample, this relation cannot be directly applied.

To quantify the effect of IMF variations on the SFR, we use the \texttt{PÉGASE} stellar population synthesis code to generate a stellar population consistent with our measured IMF slopes, assuming a constant SFR of 1 $\rm M_\odot \, \rm{yr^{-1}}$ over 100 Myr. From the emitted Lyman photons, we compute the H$\alpha$ luminosity and derive the conversion factor between luminosity and SFR for the IMF-consistent population. Applying this factor to the observed H$\alpha$ luminosity yields the IMF-corrected SFR, SFR$_{\rm{IMF}}$.

Figure~\ref{fig_SFRs} shows the comparison between the standard Salpeter-derived SFR and the IMF-consistent SFR. The average difference in logarithmic SFR, $\Delta = \log(\rm{SFR_{IMF}}) - \log(\rm{SFR_{Salpeter}})$, is $-0.6$~dex, with a standard deviation of $\sigma_\Delta = 0.445$~dex, indicating substantial variation.

Several factors would need to be considered for a more robust correction of SFRs, including metallicity (and its evolution), SFH, stellar evolution effects, and binary fractions. Given these complications, and to avoid making any strong assumption, we choose to use the H$\alpha$ luminosity directly as a robust tracer of massive stars, rather than applying a potentially uncertain SFR conversion.

\begin{figure}
    \centering
    \includegraphics[width=0.9\linewidth]{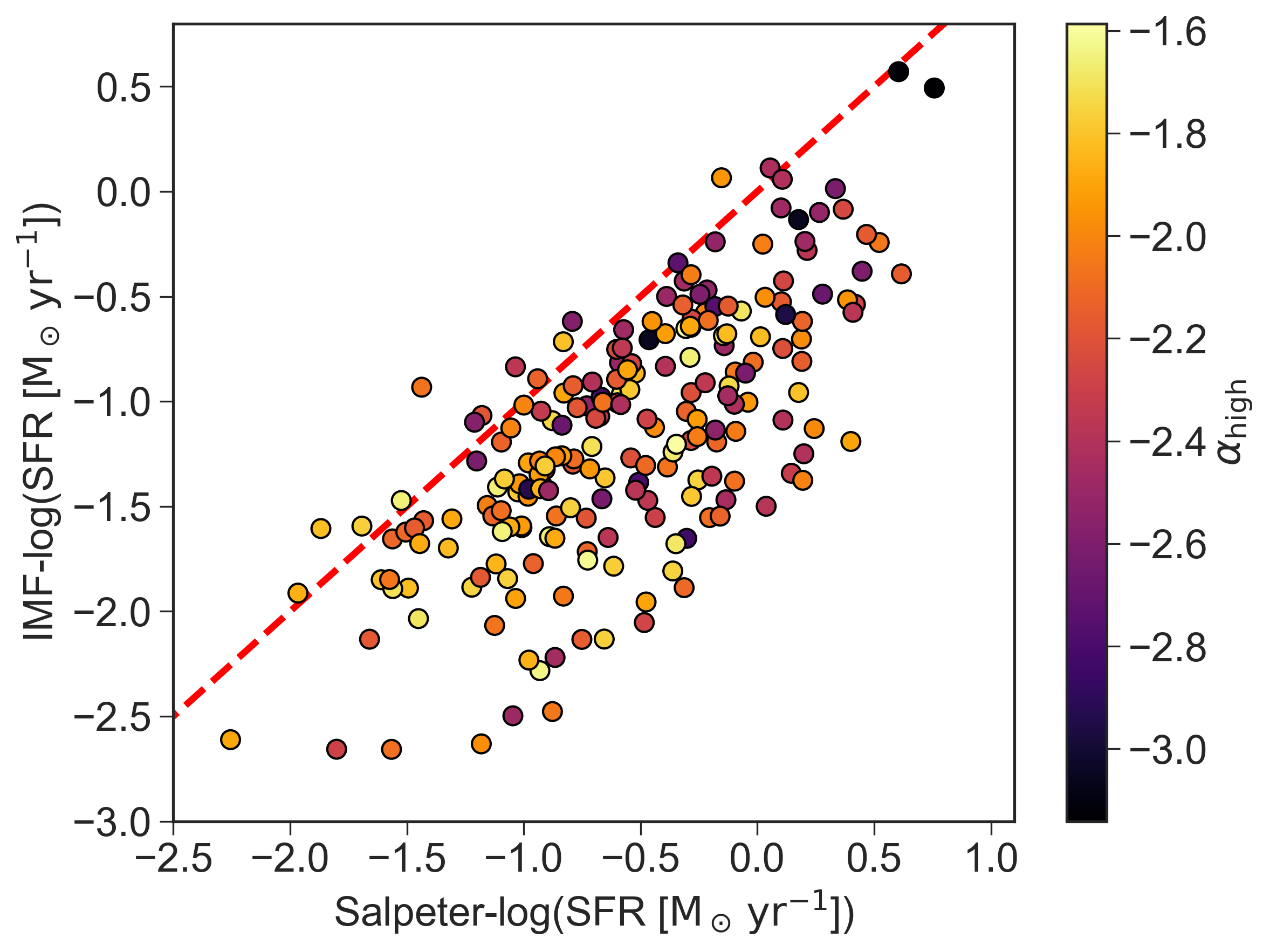}
    \caption{Comparison between logarithmic SFRs derived assuming a Salpeter IMF (x-axis) and SFRs consistent with the measured IMF (y-axis). Points are colour-coded by $\alpha_{\rm{high}}$. The red dashed line shows the one-to-one relation. The difference, $\Delta = \log(\rm{SFR_{IMF}}) - \log(\rm{SFR_{Salpeter}})$, has a mean of $-0.6$ dex and a standard deviation of 0.445 dex, suggesting that variations in the IMF can significantly impact the inferred SFR.}
    \label{fig_SFRs}
\end{figure}

\end{document}